\documentclass[12pt, notitlepage,aps,preprintnumbers,superscriptaddress,nofootinbib,aps,prd,floatfix]{revtex4-2}
\usepackage{float}
\usepackage{mathrsfs}
\usepackage{array}
\usepackage{epsfig}
\usepackage{amssymb}   
\usepackage{graphicx}   
\usepackage{verbatim}   
\usepackage{color}      
\usepackage{subfigure}  
\usepackage[colorlinks=true, linkcolor=blue, citecolor=blue, urlcolor=blue]{hyperref}

\usepackage[utf8]{inputenc}  
\usepackage{textcomp}       
\usepackage{amsmath}        

\definecolor{darkblue}{rgb}{0.0,0,0.5}
\definecolor{darkgreen}{rgb}{0.0,0.5,0}
\usepackage[normalem]{ulem}

\begin{document}
\preprint{MSUHEP-26-002}

\title{A Fast Method for Correlated Updates of Proton PDFs and the Strong Coupling $\alpha_s$}

\author{Yao Fu}
\email{fuyao3@msu.edu}
\author{Carl Schmidt}
\email{schmidt@pa.msu.edu}
\author{ C.--P. Yuan}
\email{yuan@pa.msu.edu}
\affiliation{
Department of Physics and Astronomy, Michigan State University,\\
East Lansing, MI 48824 U.S.A. }

\date{\today}
\begin{abstract}

We present an extended version of the \texttt{ePump} framework that enables the simultaneous profiling of proton parton distribution functions (PDFs) and the strong coupling $\alpha_s$ using new experimental data. By promoting $\alpha_s$ to a fit parameter within the Hessian updating formalism, the method performs coherent updates of $\{\text{PDFs},\alpha_s\}$ while preserving parameter correlations and the full covariance structure. Validation studies based on CTEQ-TEA analyses with collider data demonstrate that the upgraded \texttt{ePump} accurately reproduces the shifts in PDFs, the preferred $\alpha_s(m_Z)$, and the associated uncertainty reductions obtained in full global fits, including those inferred from Lagrange--Multiplier scans; small deviations arise only for data sets whose $\chi^2$ profiles exhibit nonlinear behavior.
Applications to representative collider measurements illustrate the impact on the gluon distribution and on precision observables such as the Higgs boson production cross section via gluon fusion. This enhanced framework provides a fast and reliable tool for assessing the effects of new data on the global QCD parameter space, offering near-global-fit accuracy at a fraction of the computational cost.

\end{abstract}

\maketitle

\section{Introduction}\label{sec:Intro}

A quantitatively precise description of hadronic collisions at the Tevatron and the Large Hadron Collider (LHC) requires two foundational QCD ingredients: the parton distribution functions (PDFs) of the proton and the strong coupling constant, $\alpha_s$. In global PDF analyses such as those carried out by the CTEQ-TEA collaboration, PDFs are obtained by fitting a flexible nonperturbative parametrization at an input scale $Q_0 \sim \mathcal{O}(1~\mathrm{GeV})$ and evolving to higher scales with the DGLAP equations. The fit combines a broad set of hard-scattering measurements and employs robust statistical treatments of correlated and uncorrelated experimental uncertainties in a Hessian framework. This approach yields a central PDF set and a set of eigenvector ``error PDFs'' that propagate PDF uncertainties to collider observables in a transparent way. $\alpha_s$ is a central parameter in these computations—entering both the perturbative partonic cross sections and the PDF evolution—and its value and uncertainty must be handled consistently with the PDFs to ensure reliable theory predictions and uncertainty budgets.

The high-precision physics program at the LHC has sharpened the need for methods that can rapidly incorporate new, high-statistics measurements into existing PDF ensembles while tracking the interplay with $\alpha_s$. A sufficiently precise knowledge of $\alpha_s(m_Z)$ is critical across Standard Model (SM) tests and new-physics searches, and the world average is now at the sub-percent level, $\alpha_s(m_Z)=0.118\pm0.0009$~\cite{ref:pdg}. New determinations from LHC and deep‑inelastic scattering (DIS) experiments reach comparable precision—for example, the ATLAS extraction from the recoil (transverse-momentum) spectrum of $Z$ bosons at $\sqrt{s}=8~\text{TeV}$ finds $\alpha_s(m_Z)=0.1183\pm0.0009$~\cite{ref:ATLAS_AlphaS}, while a recent CMS analysis that \emph{simultaneously} fits $\alpha_s$ and PDFs using inclusive jet cross sections at multiple center-of-mass energies, combined with HERA DIS data, reports $\alpha_s(m_Z)=0.1176^{+0.0014}_{-0.0016}$~\cite{ref:CMS_AlphaS}. In DIS, fits that simultaneously determine $\alpha_s$ and the proton PDFs—e.g., the H1 analysis with inclusive jet and dijet data—obtain values such as $\alpha_s(m_Z)=0.1142\pm0.0028$~\cite{ref:DIS_AlphaS}, with a direct handle on the gluon–$\alpha_s$ correlation. From the nonperturbative side, lattice QCD averages compiled by the Flavour Lattice Averaging Group (FLAG) reach a precision $\alpha_s(m_Z)=0.1183\pm0.0007$~\cite{ref:FLAG,ref:Lattice_AlphaS}, underscoring the need for comparable accuracy from phenomenological extractions and for coherent combinations across methods. Comprehensive recent reviews summarize both the current status and the opportunities ahead to further reduce the uncertainty on $\alpha_s$.

An essential challenge in this landscape is to \emph{quantify the mutual correlation} between $\alpha_s$ and PDFs when incorporating new hadron-collider data. PDF uncertainties are most commonly represented and propagated through either (i) Hessian error sets or (ii) Monte-Carlo (MC) replica ensembles. Reweighting methods in the MC-replica paradigm~\cite{ref:Reweighting_MC} have long been used to assess the impact of new data without refitting, while the Hessian updating or ``profiling'' approach~\cite{ref:Reweighting_Hessian} provides a computationally efficient alternative for Hessian PDF sets, with a well-defined treatment of global tolerances. The formal relationships between these strategies, and their respective assumptions, have been extensively studied.

The \texttt{ePump} (Error PDF Updating Method Package) program~\cite{ref:ePump,ref:ePump2} implements a general and fast Hessian profiling procedure to update a given Hessian PDF ensemble with new experimental information—returning an updated best-fit PDF and a rotated set of error PDFs, together with updated predictions for arbitrary observables. The underlying derivation expands the $\chi^2$ about the global minimum in the eigenvector basis, augments it with the covariance of the new data, and solves for the updated minimum and error eigenvectors in closed form, including options for dynamical tolerances and optimized reduced error sets. The formalism and software have been validated against full global fits and made publicly available.

In its original formulation, however, the \texttt{ePump} framework assumes a fixed value of the strong coupling constant $\alpha_s$. This assumption can be restrictive for processes that exhibit strong correlations between the PDFs and $\alpha_s$—particularly those involving the gluon distribution. Examples include Drell-Yan production, inclusive jet production, top-quark pair production, and Higgs boson production via gluon fusion. For such observables, new experimental measurements may simultaneously constrain the gluon PDF and modify the preferred value of $\alpha_s$.

In this work we present a major upgrade of \texttt{ePump} that extends the Hessian updating formalism to \emph{fit $\alpha_s$ and PDFs simultaneously} from new measurements, targeting in particular high-precision inputs from the LHC. Methodologically, $\alpha_s$ is treated as an additional degree of freedom in the Hessian parameter space, and the linearized response of the theory predictions is augmented to include derivatives with respect to $\alpha_s$, thereby updating both the PDF parameters and $\alpha_s$ in a single profiling step. This generalizes earlier profiling concepts to the joint $\{\text{PDF},\alpha_s\}$ space and parallels ideas developed in the Hessian-profiling literature. The upgraded framework is designed to preserve the rigorous error propagation of the Hessian method, account for correlated experimental systematics through full covariance matrices, and accommodate dynamical tolerances carried over from the parent global analysis. Building on earlier validations of \texttt{ePump} against full CTEQ-TEA global fits, we find that the updated \texttt{ePump} reproduces, to very good accuracy, the results of Lagrange–Multiplier scans in which $\alpha_s$ is profiled while refitting PDFs, but \emph{without} invoking the full global-fit machinery.

From a practical standpoint, the resulting method provides a computationally efficient approximation to a global QCD refit in the combined parameter space $\{\text{PDFs},\alpha_s\}$. Starting from an existing Hessian PDF set, the extended \texttt{ePump} procedure produces updated central PDFs, a corresponding shift in the preferred value of $\alpha_s$, and a rotated set of eigenvector error PDFs that encode the modified uncertainty structure. In typical applications the updating procedure requires only seconds to minutes once theoretical predictions for the new data are available, making it orders of magnitude faster than a full global PDF refit. With the increasing precision of measurements at the LHC, and with the much larger datasets anticipated from the High-Luminosity LHC, such efficiency is essential: the extended \texttt{ePump} framework provides a practical tool for rapidly exploring the implications of new collider data and for guiding future global-analysis efforts.

The remainder of the paper is organized as follows. 
In Sec.~\ref{sec:Updating}, we present the extended formalism, detailing the incorporation of $\alpha_s$ into the Hessian updating equations, the associated theoretical inputs, and the numerical implementation.
In Sec.~\ref{sec:validation} we present validation studies—comparing \texttt{ePump}'s $\{\text{PDF},\alpha_s\}$ updates against full refits and Lagrange–Multiplier scans—and quantify the agreement for representative observables. Section~\ref{sec:demo} demonstrates physics applications using recent LHC data sets most sensitive to $\alpha_s$ and to the gluon PDF. We conclude in Sec.~\ref{sec:conclusion} with a discussion of prospects for HL-LHC precision and future global-fit synergy.

\section{Updating Hessian Error PDFs and $\alpha_s$ From New Experimental Data}\label{sec:Updating}

\subsection{Review of the Hessian Method, including $\alpha_s$}\label{sec:ReviewHessian}

The PDFs\footnote{Note that we suppress the flavor index of the PDFs.  All PDF indices in
these sections will correspond to eigenvector directions.} $f(x,Q_0;{\bf z})$, defined at the initial scale $Q_0$, are parametrized by $N$ parameters $\{z_i;\ i=1,N\}$, which we collect into a vector $\bf z$.   The determination of the PDFs is obtained using a $\chi^2$-function, which quantifies the discrepancy between the theoretical predictions and the experimental measurements of a global set of experiments, including the experimental errors.  In addition to the parameters, $z_i$, the $\chi^2$-function also depends on $\alpha_s\equiv\alpha_s(M_Z)$. The best-fit PDFs are then obtained by minimizing $\chi^2$ as a function of the parameters, for a given $\alpha_s$.

The $\chi^2$ function can be written, without loss of generality and up to quadratic order in ${\bf z}$ and $\alpha_s$, as
\begin{eqnarray}
\Delta\chi^2({\bf z},z_{\alpha})&=&T^2\left[\sum_{i=1}^N(z_i-b_iz_\alpha)^2+z_\alpha^2\right]\,,\label{eq:chisquare}
\end{eqnarray}
where 
\begin{eqnarray}
z_{\alpha}&=&\frac{\alpha_s-\alpha_s^0}{\delta\alpha_s}\,,\label{eq:zalpha}
\end{eqnarray}
with $\alpha_s^0$ and $\delta\alpha_s$ the best-fit value and the uncertainty of $\alpha_s$ at the same prescribed confidence level (CL) as used for the Hessian error PDFs .  The PDF parameters $z_i$ have been defined and scaled in terms of the Hessian eigenvalue directions around the minimum of $\chi^2$ for $z_\alpha=0$, while the $b_i$ variables give correlations between the $z_i$ and $\alpha_s$.  
Note that one can use either $\alpha_s^0=\alpha_\mathrm{GA}$ and $\delta\alpha_s=\delta\alpha_\mathrm{GA}$, the best-fit value and the uncertainty of $\alpha_s$ determined purely by the global analysis (GA) data, or one can use $\alpha_s^0=\alpha_\mathrm{WA}$ and $\delta\alpha_s=\delta\alpha_\mathrm{WA}$, the world-average (WA) best-fit value and  uncertainty of $\alpha_s$. In the latter case, Eq.~(\ref{eq:chisquare}) is obtained by including a penalty term to the $\chi^2$ function, which takes into account all of the data used to constrain the world-average fitting of $\alpha_s$ that has not already been included in the global analysis, assuming that this additional data is uncorrelated with choice of the PDF parameters $z_i$.  See Appendix~\ref{sec:WorldAverage} for a more detailed discussion.

The uncertainties in the PDF parameters and $\alpha_s$ are determined by the requirement $\Delta\chi^2\le T^2$ at a prescribed confidence level (CL), where $T$ is the tolerance parameter.\footnote{For perfectly Gaussian and mutually consistent data errors, one would set $T=1$ at the 68\% CL (or equivalently $T=1.645$ at the 90\% CL). In practice, $T$ is increased to absorb experimental inconsistencies among data sets and uncertainties associated with the choice of nonperturbative parametrization. The CTEQ-TEA group has adopted $T=10$ at the 90\% CL in their CT18 analyses~\cite{ref:CT18}.}
Letting $z^\prime_i=z_i-b_iz_\alpha$ and $z^\prime_{N+1}=z_\alpha$, this yields
\begin{eqnarray}
\sum_{i=1}^{N+1}(z^\prime_i)^2\le1\label{eq:uncertaintyP}
\label{eq:hypersphere}
\end{eqnarray}
at the same CL.
This gives the standard results when we $z_\alpha=0$ is imposed. 
Here, the best-fit PDFs are given by
$f^0(x,Q_0)\equiv f(x,Q_0;\mathrm{\bf 0})$, and the $2N$ error PDFs (2 for each eigenvector direction) are defined by $f^{\pm j}(x,Q_0)\equiv f(x,Q_0;\pm{\bf e}^{j})$,
where $({\bf e}^{j})_i=\delta^j_i$.  For the combind PDF$+\alpha_s$ uncertainties, we add two additional PDFs
$f^{\pm \alpha}(x,Q_0)\equiv f(x,Q_0;\pm{\bf b})$, which correspond to the best-fit PDFs for $z_\alpha=\pm1$ (or equivalently $\alpha_s=\alpha_s^0\pm\delta\alpha_s$).

To find the combined PDF$+\alpha_s$ uncertainty on a theoretical prediction for some observable $X$ in the Hessian approximation, we write it as a function of the PDF parameters and $\alpha_s$, and expand to linear order in a Taylor expansion around the best fit,
\begin{eqnarray}
X({\bf z},z_\alpha)&=&X({\bf z}^\prime+{\bf b}z_\alpha,z_\alpha)
\nonumber\\
&=&X(\mathrm{\bf 0},0)+\sum_{j=1}^N\frac{\partial X}{\partial z_{j}}\Bigg|_{{\bf 0},0}z^\prime_j+\left(\sum_{j=1}^Nb_j\frac{\partial X}{\partial z_{j}}\Bigg|_{{\bf 0},0}+\frac{\partial X}{\partial z_{\alpha}}\Bigg|_{{\bf 0},0}\right)z_\alpha+\cdots\,
 .\label{eq:observable}
\end{eqnarray}
We can calculate $X(\mathrm{\bf 0},0)\equiv X(f^0,\alpha_s^0)$ using the best-fit PDFs. The first derivatives with respect to the PDF parameters can be calculated numerically using the
$+$ and $-$ error PDFs in each eigenvector direction $j$ as
\begin{eqnarray}
\frac{\partial X}{\partial z_{j}}\Bigg|_{{\bf 0},0}\,\approx\,\Delta X^{j}&=&
\frac{X(+{\bf e}^j,0)-X(-{\bf e}^j,0,)}{2}\nonumber\\
&\equiv&
\frac{X(f^{+j},\alpha_s^0)-X(f^{-j},\alpha_s^0)}{2}\,.\label{eq:dx}
\end{eqnarray}
Finally, the last term in parentheses can be calculated numerically using the $f^{\pm\alpha}$ PDF sets as
\begin{eqnarray}
\left(\sum_{j=1}^Nb_j\frac{\partial X}{\partial z_{j}}\Bigg|_{{\bf 0},0}+\frac{\partial X}{\partial z_{\alpha}}\Bigg|_{{\bf 0},0}\right)\,\approx\,\Delta X^\alpha&=&
\frac{X(+{\bf b},+1)-X(-{\bf b},-1)}{2}
\nonumber\\
&\equiv&
\frac{X(f^{+\alpha},\alpha_s^0+\delta\alpha_s)-X(f^{-\alpha},\alpha_s^0-\delta\alpha_s)}{2}\,.\label{eq:dxalpha}
\end{eqnarray}
Thus, letting $\Delta X^{N+1}\equiv\Delta X^\alpha$ and $z^\prime_{N+1}\equiv z_\alpha$, we can write
\begin{eqnarray}
X({\bf z},z_\alpha)
&=&X(\mathrm{\bf 0},0)+\sum_{j=1}^{N+1}\Delta X^jz^\prime_j+\cdots\,
 .\label{eq:observableii}
\end{eqnarray}

The combined PDF$+\alpha_s$ theoretical uncertainty in $X$ at the specified CL is determined by the maximum and minimum values of $X$,
subject to the constraint $\sum (z^\prime_i)^2\le1$.  In the quadratic approximation of Eq.~(\ref{eq:chisquare}) and the linear approximation of Eq.~(\ref{eq:observableii}),  we obtain 
\begin{eqnarray}
X^\pm&=&X(\mathrm{\bf 0})\pm \Delta X\,,\label{eq:limits}
\end{eqnarray}
where
\begin{eqnarray}
\Delta X&=&\sqrt{\sum_{j=1}^{N+1}\Bigl(\Delta X^j\Bigr)^2}\,.\label{eq:uncertainty}
\end{eqnarray}
Equation~\eqref{eq:uncertainty} is the Symmetric Master Equation for the combined Hessian PDF$+\alpha_s$ uncertainty on the observable $X$.

\subsection{Updating of Hessian error PDFs}\label{sec:UpdatePDFs}

We now address the impact of new data on the PDFs and $\alpha_s$, and on their associated uncertainties.
Suppose that we have measurements for $N_X$ observables with experimental values given by $X^E_\alpha$, as well as the inverse covariance matrix $C^{-1}_{\alpha\beta}$ for the correlated experimental errors in the measurements.  The contribution of these new data in the global $\chi^2$ function is
\begin{eqnarray}
\Delta\chi^2({\bf z},z_\alpha)_{\rm new}&=&T^2\sum_{i=1}^{N+1}(z^\prime_i)^2\,+\,\sum_{\alpha,\beta=1}^{N_X}\left(X_\alpha({\bf z},z_\alpha)-X_{\alpha}^E\right)C^{-1}_{\alpha\beta}\left(X_\beta({\bf z},z_\alpha)-X_{\beta}^E\right)\,,\label{eq:chi2new}
\end{eqnarray}
where we again let $z^\prime_i=z_i-b_iz_\alpha$ and $z^\prime_{N+1}=z_\alpha$.
Using the generalization of Eqs.~(\ref{eq:dx}), (\ref{eq:dxalpha}), and (\ref{eq:observableii}) to expand $X_\alpha({\bf z},z_\alpha)$ to linear order in ${\bf z}^\prime$, gives
\begin{eqnarray}
\Delta\chi^2({\bf z},z_\alpha)_{\rm new}&=&
\sum_{\alpha,\beta=1}^{N_X}\left(X_\alpha(\mathrm{\bf 0},0)-X^E_\alpha\right)C^{-1}_{\alpha\beta}\left(X_\beta(\mathrm{\bf 0},0)-X^E_\beta\right)\nonumber\\
 &&+\,T^2\left[\sum_{i=1}^{N+1}(z^\prime_i)^2\,+\,\sum_{i,j=1}^{N+1}z^\prime_iM^{ij}z^\prime_j\,-\,2\sum_{i=1}^{N+1}z^\prime_iA^i\right]\,,\label{eq:chi2newPZ}
\end{eqnarray}
where
\begin{eqnarray}
A^i&=&
\frac{1}{T^2}\sum_{\alpha,\beta=1}^{N_X}\left(X^E_\alpha-X_\alpha(\mathrm{\bf 0},0)\right)\,C^{-1}_{\alpha\beta}\,\Delta{X}^{i}_{\beta}\,,\nonumber\\
M^{ij} &=&\frac{1}{T^2}\sum_{\alpha,\beta=1}^{N_X}\Delta{X}^{i}_{\alpha}\,C^{-1}_{\alpha\beta}\,\Delta{X}^{j}_{\beta}\ .\label{eq:AandM}
\end{eqnarray}

With the additional data added to the $\chi^2$ function, the new best-fit parameters are
\begin{eqnarray}
\bar{z}_i^{\,\prime}&=&\sum_{j=1}^{N+1}(\delta+M)^{-1}_{ij}A^j\, .\label{eq:bestfit}
\end{eqnarray}
Setting $\bar{z}_{N+1}^{\,\prime}=\bar{z}_\alpha$, the updated best-fit value of $\alpha_s$ is given by
\begin{eqnarray}
\alpha^0_{s,\mathrm{new}}&=&\alpha_s^0+\bar{z}_\alpha\,\delta\alpha_s \, .\label{eq:bestfitalpha}
\end{eqnarray}
Using the generalization of Eqs.~(\ref{eq:dx}), (\ref{eq:dxalpha}), and (\ref{eq:observableii}) to expand $f(x,Q_0,{\bf z})$ to linear order in ${\bf z}^\prime$, we can approximate the new best-fit PDFs, corresponding to the new best-fit $\alpha^0_{s,\mathrm{new}}$, by
\begin{eqnarray}
f^0(x,Q_0)_{\rm new}&=&f^0(x,Q_0)+\sum_{i=1}^{N+1}\bar{z}_i^{\,\prime}\,\Delta{f}^i(x,Q_0)\, ,\label{eq:bestfitPDF}
\end{eqnarray}
where
\begin{eqnarray}
\Delta f^i(x,Q_0)&=&\frac{f^{+i}(x,Q_0)-f^{-i}(x,Q_0)}{2}\ \qquad i=1,\dots,N\nonumber\\
\Delta f^{N+1}(x,Q_0)&=&\frac{f^{+\alpha}(x,Q_0)-f^{-\alpha}(x,Q_0)}{2}\,.
\label{eq:DF}
\end{eqnarray}

Now to determine the updated uncertainties on the PDFs and $\alpha_s$.  Here, we treat $z^\prime_i$ and $z_\alpha$ differently, because in determining the uncertainty in $\alpha_s$, the $z^\prime_i$ are always adjusted to minimize $\chi^2$ for a given $z_\alpha$.
Up to an irrelevant constant, we can write the updated $\chi^2$ function as
\begin{eqnarray}
\Delta\chi^2({\bf z},z_\alpha)_{\rm new}&=&T^2\sum_{i,j=1}^{N+1}(z^\prime_i-\bar{z}^{\,\prime}_i)(\delta+M)^{ij}(z^\prime_j-\bar{z}^{\,\prime}_j)\nonumber\\
&=&T^2\left[\left(1+M^{N+1,N+1}\right)(z_\alpha-\bar{z}_\alpha)^2+2\sum_{i=1}^{N}(z_\alpha-\bar{z}_\alpha)M^{N+1,i}(z^\prime_i-\bar{z}^{\,\prime}_i)\right.\nonumber\\
&&\left.
+\sum_{i,j=1}^{N}(z^\prime_i-\bar{z}^{\,\prime}_i)(\delta+M)^{ij}(z^\prime_j-\bar{z}^{\,\prime}_j)\right]\\
&=&T^2\left[\left(\frac{\delta\alpha_s}{\delta\alpha_{s,\mathrm{new}}}\right)^2(z_\alpha-\bar{z}_\alpha)^2
+\sum_{i,j=1}^{N}(z^\prime_i-\bar{z}^{\,\prime}_i-d_i)(\delta+M)^{ij}(z^\prime_j-\bar{z}^{\,\prime}_j-d_j)\right]\,.\nonumber\label{eq:chi2newPZii}
\end{eqnarray}
where the new uncertainty in $\alpha_s$ is given by
\begin{eqnarray}
\delta\alpha_{s,\mathrm{new}}&=&\frac{\delta\alpha_s}{\sqrt{1+M^{N+1,N+1}-\sum_{i,j=1}^NM^{N+1,i}(\widehat{\delta+M})^{-1}_{ij}M^{N+1,j}}}\,,\label{eq:newdeltaalphas}
\end{eqnarray}
and
\begin{eqnarray}
d_i=-\sum_{j=1}^N(\widehat{\delta+M})^{-1}_{ij}M^{N+1,j}(z_\alpha-\bar{z}_\alpha)\,.\label{eq:deltai}
\end{eqnarray}
Here, $(\widehat{\delta+M})^{-1}_{ij}$ denotes the inverse of the $N\times N$ submatrix of $(\delta+M)^{ij}$.

The new $\alpha_s$ PDFs, $f^{\pm\alpha}$, are calculated at $\alpha^0_{s,\mathrm{new}}\pm\delta\alpha_{s,\mathrm{new}}$, with the PDF parameters chosen to minimize the $\chi^2$ function at the value of $\alpha_s$.  This corresponds to the parameters at the values 
\begin{eqnarray}
z^{(\pm)}_\alpha&=&\bar{z}_\alpha\pm\left(\frac{\delta\alpha_{s,\mathrm{new}}}{\delta\alpha_{s}}\right)\nonumber\\
z^{(\pm)\prime}_i&=&\bar{z}^{\,\prime}_i+d^{(\pm)}_i\nonumber\,,
\end{eqnarray}
with
\begin{eqnarray}
d^{(\pm)}_i&=&\mp\left(\frac{\delta\alpha_{s,\mathrm{new}}}{\delta\alpha_{s}}\right)\sum_{j=1}^N(\widehat{\delta+M})^{-1}_{ij}M^{N+1,i}\,.
\end{eqnarray}
Working at linear order in the parameters,  we obtain the new $\alpha_s$ PDFs to be

\begin{eqnarray}
f^{\pm\alpha}(x,Q_0)_\mathrm{new}&=&f^0(x,Q_0)_{\rm new}\nonumber\\
&&\pm\left(\frac{\delta\alpha_{s,\mathrm{new}}}{\delta\alpha_{s}}\right)\left[\Delta{f}^\alpha(x,Q_0)-\sum_{i,j=1}^NM^{N+1,i}(\widehat{\delta+M})^{-1}_{ij}\,\Delta{f}^j(x,Q_0)\right]\, .\label{eq:newev_alphas}
\end{eqnarray}

To obtain the updated Hessian eigenvector PDFs at the new best-fit value of $\alpha_s$, we must diagonalize the $N\times N$ matrix $M^{ij}$.
The normalized eigenvectors $U_i^{(r)}$ and eigenvalues $\lambda^{(r)}$ of this matrix satisfy
\begin{eqnarray}
\sum_{j=1}^N M^{ij}U^{(r)}_j&=&\lambda^{(r)} U_i^{(r)}\,,\nonumber\\
\sum_{i=1}^NU_i^{(r)}U_i^{(s)}&=&\delta_{rs}\,.
\end{eqnarray}
We can now simplify the equation for $\Delta\chi^2$ by introducing new coordinates ${\bf c}\equiv(c_1,\dots,c_N)$ and $c_\alpha$
defined by
\begin{eqnarray}\label{eq:zi_alpha}
z^\prime_i&=&\bar{z}^{\,\prime}_i+\sum_{r=1}^N\frac{1}{\sqrt{1+\lambda^{(r)}}}\,c_r\,U_i^{(r)}\nonumber\\
z_\alpha&=&\bar{z}_\alpha+\left(\frac{\delta\alpha_{s,\mathrm{new}}}{\delta\alpha_{s}}\right)c_\alpha\,.
\end{eqnarray}
In terms of these variables, we obtain 
\begin{eqnarray}
\Delta\chi^2({\bf z},z_\alpha)_{\rm new}&=&T^2\left[\sum_{r=1}^N(c_r-d_rc_\alpha)^2+c_\alpha^2\right]\,,\label{eq:chi2newPZdiagonalized}
\end{eqnarray}
where
\begin{eqnarray}\label{eq:newdistance}
d_r=-\frac{\left(\delta\alpha_{s,\mathrm{new}}/\delta\alpha_{s}\right)}{\sqrt{1+\lambda^{(r)}}}\sum_{i=1}^NM^{N+1,i}U_i^{(r)}\,.
\end{eqnarray}
The updated error PDFs are obtained by setting the new coordinates to ${\bf c}^{\pm(r)}=\pm{\bf e}^r$ with $c_\alpha=0$ ({\it i.e.,} evaluated at $\alpha_s=\alpha_{s,\mathrm{new}}^0$).  Working at linear order in the parameters we obtain

\begin{eqnarray}
f^{\pm(r)}(x,Q_0)&=&f^0(x,Q_0)_{\rm new}+\frac{1}{\sqrt{1+\lambda^{(r)}}}  \sum_{i=1}^NU_i^{(r)}\,\Delta{f}^i(x,Q_0)\, .\label{eq:newev}
\end{eqnarray}
It is worth noting that the size of the updated uncertainty bands from the Hessian error PDFs  is identical to the result obtained when $\alpha_s$ dependence is neglected in the updating.  
The central value of the PDFs is shifted by the change in $\alpha_s$, but the curvature of $\chi^2$ with respect to the PDF parameters—and hence the error-PDF spread—is unaffected. This follows directly from the quadratic approximation: at quadratic order, a change in $\alpha_s$ can shift the minimum of $\chi^2$ as a function of $z_i$, but it cannot alter the curvature.

\subsection{Practical Considerations in the Extended \texttt{ePump} Implementation}
\label{sec:Practical}

The two preceding subsections developed the formalism for the fixed-tolerance case, in
which the same tolerance parameter $T$ governs all Hessian eigenvector directions. In
practice, however, the Hessian eigenvector PDFs are often evaluated with
\textit{dynamical tolerances}, so that the $\pm$ error PDFs in eigenvector direction $j$
correspond to $\Delta\chi^{2} = (T^{\pm}_{j})^{2}$, which can differ both from $T^{2}$ and
from one direction to another. Similarly, the world–average value of $\alpha_{s}$
($\alpha_{s}^{\mathrm{WA}}$) and its uncertainty ($\delta\alpha_{s}^{\mathrm{WA}}$) generally
differ from the corresponding quantities determined solely by the global–analysis data
($\alpha_{s}^{\mathrm{GA}}, \delta\alpha_{s}^{\mathrm{GA}}$). These two practical considerations—
dynamical tolerances and the choice of $\alpha_{s}$ reference values—must be accounted for
in the \texttt{ePump} implementation, and their treatment is detailed in Appendices~\ref{sec:Dynamical Tolerances} and~\ref{sec:WorldAverage},
respectively. A brief summary of each point is given below.

\subsubsection*{Dynamical tolerances}

In the presence of dynamical tolerances, the $\pm$ Hessian PDFs in eigenvector direction
$j$ are evaluated at parameter vectors
\[
z = \pm \frac{T^{\pm}_{j}}{T} \, e_{j}.
\]
As shown in Appendix~\ref{sec:Dynamical Tolerances}, the principal change is to replace the symmetric first--derivative
step $\Delta X_{j}$ by its dynamically weighted counterpart $\widehat{\Delta X}_{j}$
(Eq.~(\ref{eq:dxalphaDyn})), and similarly for the PDF first differences $\widehat{\Delta f}_{j}$. All
subsequent formulae for the updated best-fit PDFs in Sec.~\ref{sec:UpdatePDFs} then carry over verbatim under the replacement
$\Delta X_{j}\rightarrow \widehat{\Delta X}_{j}$ and $\Delta f_{j}\rightarrow \widehat{\Delta f}_{j}$.
In particular, the tolerance parameter $T^{2}$ cancels from every physical observable, and the final results depend only on the dynamical tolerances, $T^{\pm}_{j}$.   
The formula for $\alpha_{s,\mathrm{new}}^{0}$ (Eq.~(\ref{eq:bestfitalphadt})) and for the
updated $\alpha_{s}$ PDFs $f^{\pm}_{\alpha}(x, Q_{0})_{\mathrm{new}}$ (Eq.~(\ref{eq:newev_dyn})) include explicit
factors of $(T_{\alpha}/T)$ that encode the dynamical tolerance assigned to the
$\alpha_{s}$ direction (again, with the factor of $T$ cancelling out of the final result.) As for the treatment of the updated Hessian eigenvector PDFs at the new best-fit value of $\alpha_s$ when using dynamical tolerances, this is identical to that of the fixed-$\alpha_s$ treatment described in Ref.~\cite{ref:ePump}.

\subsubsection*{Choice of $\alpha_{s}$ reference values}

As discussed in Appendix~\ref{sec:WorldAverage}, two natural choices exist for the reference values
$(\alpha_{s}^{0}, \delta\alpha_{s})$ entering the updating formulae: the global–analysis values
$(\alpha_{s}^{\mathrm{GA}},\delta\alpha_{s}^{\mathrm{GA}})$ and the world–average values
$(\alpha_{s}^{\mathrm{WA}},\delta\alpha_{s}^{\mathrm{WA}})$. 
When working in the quadratic approximation for $\Delta\chi^2$, the correlation between
$\alpha_s$ and the PDF parameters does not depend on the particular value of $\alpha_s$ that we expand around. Thus, when the error PDFs are provided at fixed
$\alpha_{s}=\alpha_{s}^{\mathrm{WA}}$ and $\alpha_{s}^{\mathrm{WA}}\pm \delta\alpha_{s}^{\mathrm{WA}}$ (as is conventional in most Hessian global analyses),  
the change from the updating analysis with $(\alpha_{s}^{\mathrm{GA}},\delta\alpha_{s}^{\mathrm{GA}})$ can be absorbed into a simple reparametrisation of the $\alpha_{s}$ and PDF coordinates (Eq.~(\ref{eq:chisquarebariii}-\ref{eq:barpars})).
The net effect is that the combined PDF+$\alpha_{s}$ updating analysis is formally identical
in both cases; only the numerical values of $\alpha_{s}^{0}$, $\delta\alpha_{s}$, must be adjusted accordingly. 
In practice, \texttt{ePump} accepts either convention as input, and returns updated results determined with respect to these input values.

\subsubsection*{Lagrange–Multiplier Scans}

A particularly useful capability of the extended \texttt{ePump} framework is the rapid
evaluation of Lagrange–Multiplier (LM) scans for arbitrary observables, including
$\alpha_{s}$ itself. As derived in Appendix~\ref{sec:LMScan}, the LM constraint term is additive in the
$\chi^{2}$ function and enters the updating equations through modified vectors $A_{i}$ and a
zero contribution to $M_{ij}$ (Eq.~(\ref{eq:AM_LMScan})). Because the updating matrix is assembled from
the same ingredients used for the central–value profiling, the LM scan requires no
additional theory calculations beyond those already performed for the profiling step.
A scan over $\alpha_{s}$ is obtained by treating $\alpha_{s}$ as the constrained observable:
for each target value $\alpha_{s,c}$, one applies the standard \texttt{ePump} update with a
Lagrange–Multiplier term that fixes the mean value of the profiled $\alpha_{s}$ to
$\alpha_{s,c}$. The resulting $\chi^{2}$ profile as a function of $\alpha_{s}$ can then be
compared directly with full global–analysis LM scans, as demonstrated in Sec.~\ref{sec:LMScan_practice}.

\subsubsection*{Implementation summary}

The extended \texttt{ePump} code implements the above steps as follows. Given a Hessian PDF
set $\{ f^{0}, f^{\pm j}, f^{\pm\alpha} \}$ and a new data set with experimental values
$X^{E}_{\alpha}$ and inverse covariance matrix $C^{-1}_{\alpha\beta}$, together with the
corresponding theory predictions computed at each error PDF, \texttt{ePump} proceeds in four
stages:
\begin{enumerate}
	\item Assemble the sensitivity vectors $A_{i}$ and the sensitivity matrix $M_{ij}$
	from Eq.~(\ref{eq:AandM}), using the dynamically weighted differences where appropriate.
	\item Solve the linear system of Eq.~(\ref{eq:bestfit}) to obtain the updated parameters
	$\bar{z}'_{i}$.
	\item Construct the updated central PDF $f^{0}_{\mathrm{new}}$ (Eq.~(\ref{eq:bestfitPDF})), the updated
	$\alpha_{s,\mathrm{new}}^{0}$ (Eq.~(\ref{eq:bestfitalpha}) or (\ref{eq:bestfitalphadt})), and the updated uncertainty
	$\delta\alpha_{s,\mathrm{new}}$ (Eq.~(\ref{eq:newdeltaalphas})).
	\item Diagonalize the updated $N\times N$ sensitivity matrix to obtain the new Hessian
	eigenvectors and the corresponding error PDFs (Eqs.~(\ref{eq:zi_alpha}), (\ref{eq:chi2newPZdiagonalized}), (\ref{eq:newdistance}), (\ref{eq:newev})).
\end{enumerate}
The full procedure typically requires only seconds once the theory predictions are
available, enabling rapid exploratory studies that would otherwise demand a complete
global refit.

\section{Validation of the Extended \texttt{ePump} Framework}\label{sec:validation}

A central requirement for the upgraded  \texttt{ePump}  package—now capable of simultaneously profiling the PDFs and the strong coupling constant $\alpha_s$—is to demonstrate that the Hessian‑updating formalism accurately reproduces the results of full global QCD analyses when new experimental constraints are incorporated. In this section we present a series of validation studies designed to probe both the numerical stability and the physics fidelity of the extended framework. Our strategy mirrors the validation benchmarks originally performed for the baseline  \texttt{ePump}  release, which compared updated PDF sets against full CTEQ‑TEA global fits using a variety of DIS, Drell–Yan, jet, and $t\bar{t}$ data sets.

\subsection{Benchmark Strategy and Data Sets}

Our benchmark strategy builds upon the CT18A NNLO global analysis~\cite{ref:CT18}, which includes the ATLAS 7~TeV $W/Z$ production data recorded in 2011 with an integrated luminosity of 4.6~fb$^{-1}$~\cite{ref:E248}. This measurement is designated as the E248 data set, following the experimental dataset numbering convention of CT18~\cite{ref:CT18}.
A modified baseline PDF ensemble, denoted CT18Am248, is then constructed by removing E248 from the CT18A fit and repeating the $\alpha_s$ scan. This ensemble serves as the common starting point for all subsequent \texttt{ePump} updates.

As a first closure test, E248 is reintroduced into CT18Am248 via \texttt{ePump}, and the resulting PDFs and $\alpha_s$ are compared with those of the full CT18A analysis. This tests whether \texttt{ePump} can faithfully reproduce the $\alpha_s$ dependence obtained in the global fit.

A second benchmark tests the CT18 analysis using the E268 data set, which corresponds to the ATLAS 7~TeV $W/Z$ production measurement collected in 2010 with an integrated luminosity of 35~pb$^{-1}$~\cite{ref:E268}. Starting again from CT18Am248, the PDFs and $\alpha_s$ are updated with E268 via \texttt{ePump} and compared with the full CT18 NNLO analysis. This comparison is particularly clean because CT18 and CT18A differ only in their choice of ATLAS 7~TeV $W/Z$ data set—CT18 uses E268 and CT18A uses E248—with all other experimental inputs held fixed.

\subsection{Comparison with Full Global Re‑Analyses}

For each update, the following quantities are examined:

\begin{itemize}
\item The updated best-fit value of $\alpha_s(m_Z)$,
\item Shifts in key PDF flavors, in particular the gluon and light-quark combinations,
\item Shifts in the Hessian eigenvectors and the corresponding uncertainty changes,
\item Predictions for selected validation observables computed with the updated PDF sets.
\end{itemize}

The results of these updates are compared directly with those obtained from the corresponding full global re-analyses. In particular, the CT18Am248 ensemble updated with the E248 (E268) data set using \texttt{ePump} is compared with the full CT18A (CT18) global fit, respectively.
Table~\ref{tab:alphas} summarizes the central values and uncertainties of $\alpha_s$ before and after the \texttt{ePump} updates. 
Because the E248 and E268 measurements have relatively weak sensitivity to $\alpha_s$, the resulting shifts in both the central value and its uncertainty are modest.
Overall, we find agreement within uncertainty between the upgraded \texttt{ePump} results and those from the full global analyses. This demonstrates that \texttt{ePump} provides a reliable and efficient approximation for assessing the correlated impact of new data on PDFs and $\alpha_s$.

To further test the sensitivity of the \texttt{ePump} updating to datasets with stronger impact on $\alpha_s$, we perform an additional benchmark study by removing two datasets simultaneously from the baseline global fit. In addition to the E248 measurement, the E544 data set (ATLAS 7~TeV, with an integrated luminosity of 4.5~fb$^{-1}$; 
inclusive jet cross sections, with  $R=0.6$~\cite{ref:E544}) is also removed. The jet data provide a stronger constraint on $\alpha_s$ through their direct dependence on the gluon-initiated subprocesses and the overall normalization of QCD cross sections. 
Upon removing both E248 and E544 from the global analysis, the preferred value of $\alpha_s$ shifts from $\alpha_s(m_Z) = 0.1166$ to $\alpha_s(m_Z) = 0.1161$.
Starting from this modified baseline, the two datasets are then added back simultaneously using the \texttt{ePump} framework with correlated PDF+$\alpha_s$ updating. After the update, the value of the strong coupling becomes $\alpha_s(m_Z) = 0.1164$, which shows a clear recovery toward the original global-fit result. The result is also summarized in table~\ref{tab:alphas}.
This study demonstrates that the \texttt{ePump} framework correctly captures the impact of datasets with significant $\alpha_s$ sensitivity and can reliably reproduce the corresponding shifts when such constraints are reintroduced. 
This additional study is intended purely as a sensitivity check.
In the remainder of this work, we use the standard CT18Am248 baseline (with only E248 removed); the E544 jet data is retained as in the original CT18A analysis.
These results are consistent with earlier ePump validation studies—conducted in the PDF-only context—in which the Hessian profiling method was found to reproduce full refits across multiple classes of data.
The present study confirms that this robustness extends naturally to the enlarged $\{\text{PDF},\alpha_s\}$ parameter space. 

\begin{table*}[hbt]
    \centering
    \begin{tabular}{c|c|c}
        \hline \hline
        Data sets added by \texttt{ePump} & $\alpha_s$ before and after update & $\alpha_s$ from global fit \\
        \hline
        E268~\cite{ref:E268} & 0.1164 $\pm$ 0.0025 $\rightarrow$ 0.1164 $\pm$ 0.0025 & CT18: 0.1164 $\pm$ 0.0025 \\
        E248~\cite{ref:E248} & 0.1164 $\pm$ 0.0025 $\rightarrow$ 0.1165 $\pm$ 0.0025 & CT18A: 0.1166 $\pm$ 0.0025 \\
        E248~\cite{ref:E248} + E544~\cite{ref:E544} & 0.1161 $\pm$ 0.0027 $\rightarrow$ 0.1164 $\pm$ 0.0025 & CT18A: 0.1166 $\pm$ 0.0025 \\
        \hline \hline
    \end{tabular}
    \caption{The best-fit values of the strong coupling constant $\alpha_s(m_Z)$ and their 68\% CL uncertainties before and after \texttt{ePump} updating with different input data sets.}
    \label{tab:alphas}
\end{table*}

\subsection{Reproduction of Lagrange–Multiplier Scans}\label{sec:LMScan_practice}

To stress-test the formalism, we reproduce Lagrange Multiplier (LM)–style scans of $\alpha_s$ in which the global analysis is repeated many times while fixing $\alpha_s$ at different values and refitting the PDFs each time. LM scans are considered the gold standard for joint PDF–$\alpha_s$ profiling due to their minimal reliance on linear approximations.

Using the upgraded  \texttt{ePump}  machinery, we perform approximate LM scans by repeatedly profiling with $\alpha_s$ shifted in small increments. The resulting $\chi^2(\alpha_s)$ curves show close agreement with those obtained from full LM scans in earlier CTEQ‑TEA studies. In particular, the curvature of the $\chi^2$ profile, which governs the extracted uncertainty on $\alpha_s$, is well reproduced by the profiling procedure—even though the latter requires only a single update per $\alpha_s$ value and no full global fit.  
This is consistent with earlier findings that \texttt{ePump} accurately captures second-order variations in the vicinity of the global best-fit solution.
See Appendix~\ref{sec:LMScan} for the detailed discussion about the Lagrange-Multiplier scans using \texttt{ePump}.

\begin{figure}[H]
    \centering
    \includegraphics[width=0.48\linewidth]{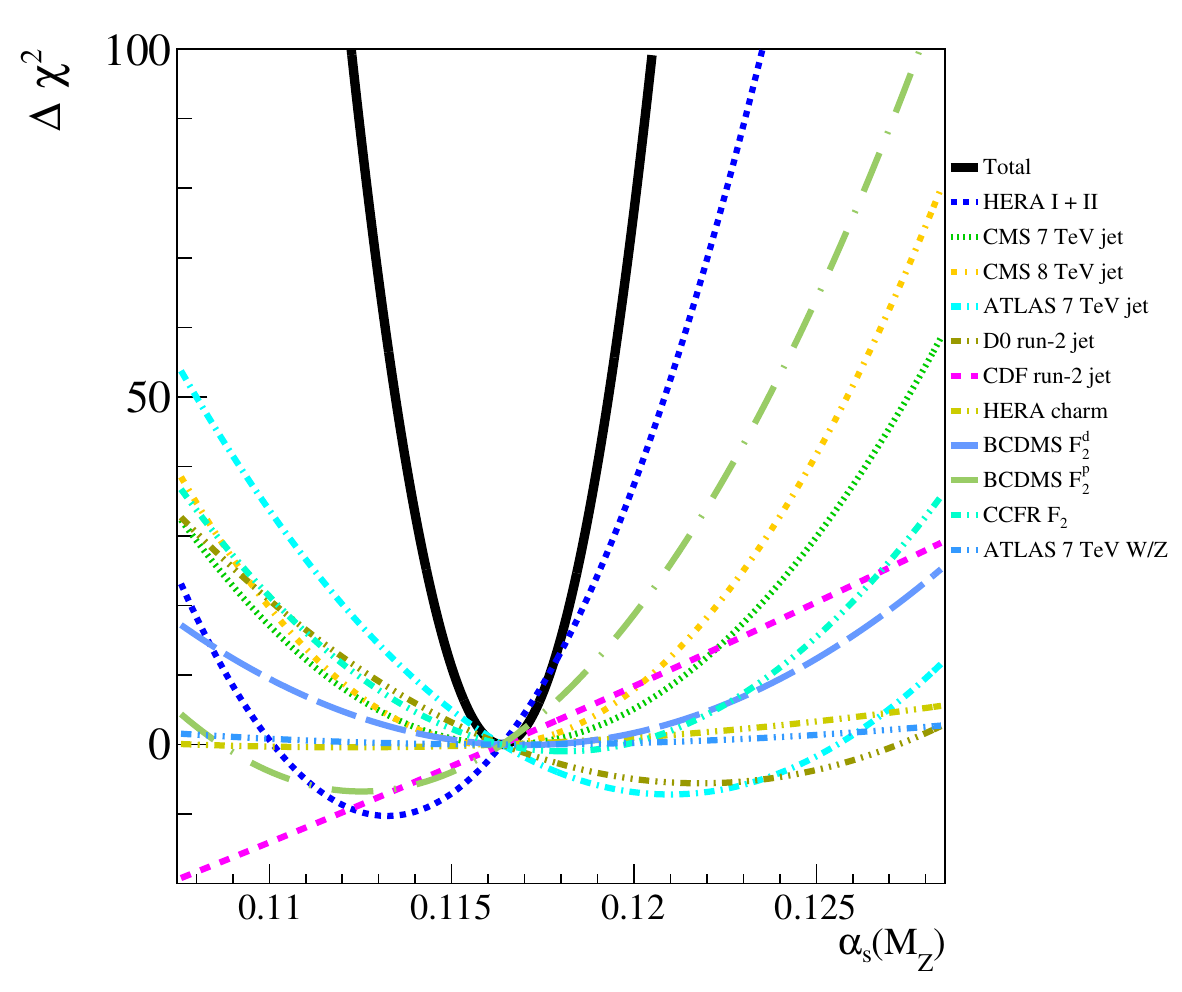}
    \includegraphics[width=0.48\linewidth]{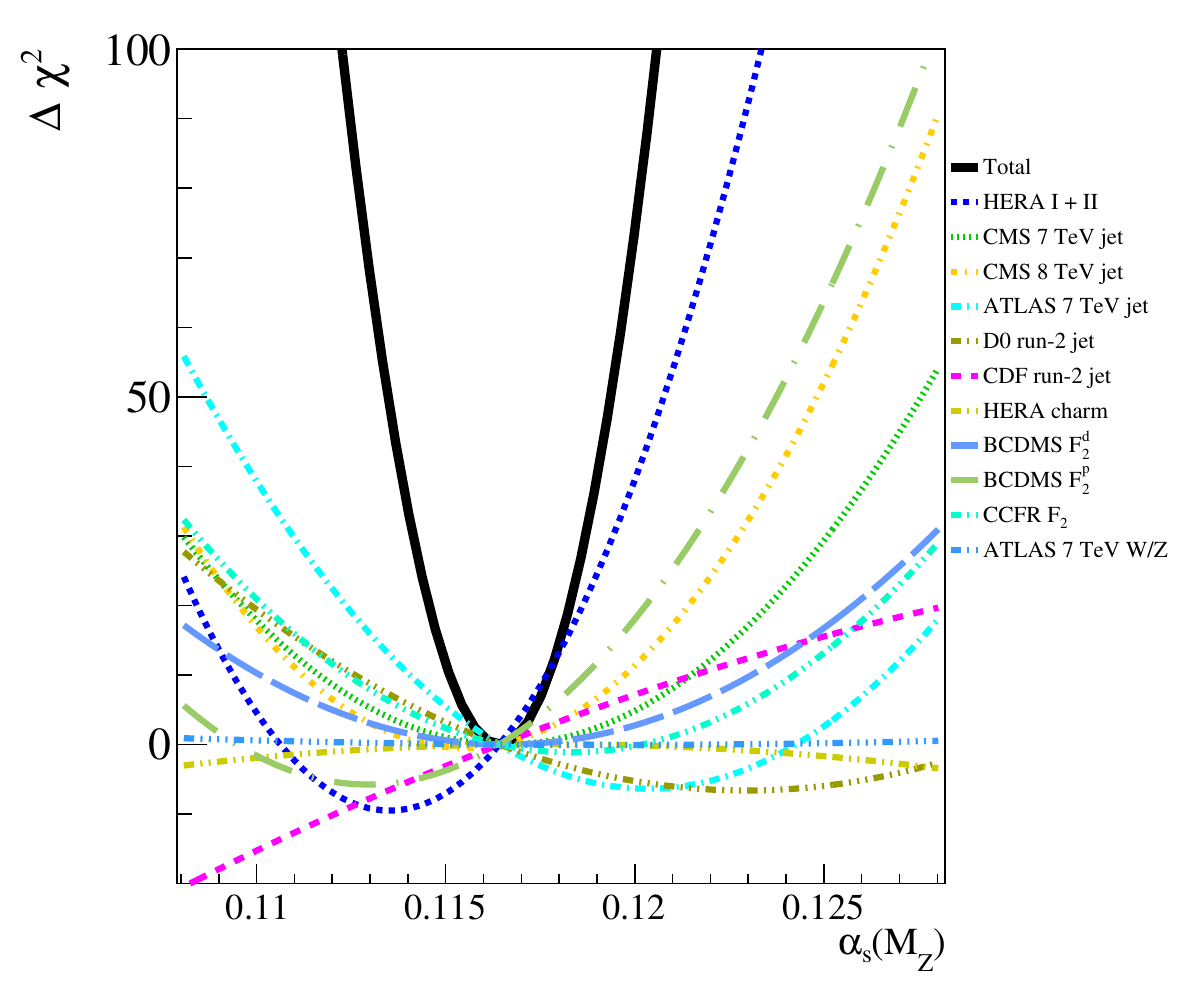}
    \caption{The scan of the strong coupling constant at the scale $M_Z$ for NNLO PDFs. Left: The baseline PDFs are obtained from the standard CT18 analysis but with the E268 data set (ATLAS 7 TeV $W/Z$, with an integrated luminosity of 35~pb$^{-1}$ collected in 2010~\cite{ref:E268}) removed, and subsequently updated by \texttt{ePump} to include the missing E268 data. The $\chi^2$ values are computed using \texttt{ePump}. Right: The corresponding results obtained from the full CT18 analysis, using the CTEQ-TEA global analysis code.}
    \label{fig:AlphaS_scan_268}
\end{figure}

\begin{figure}[H]
    \centering
    \includegraphics[width=0.48\linewidth]{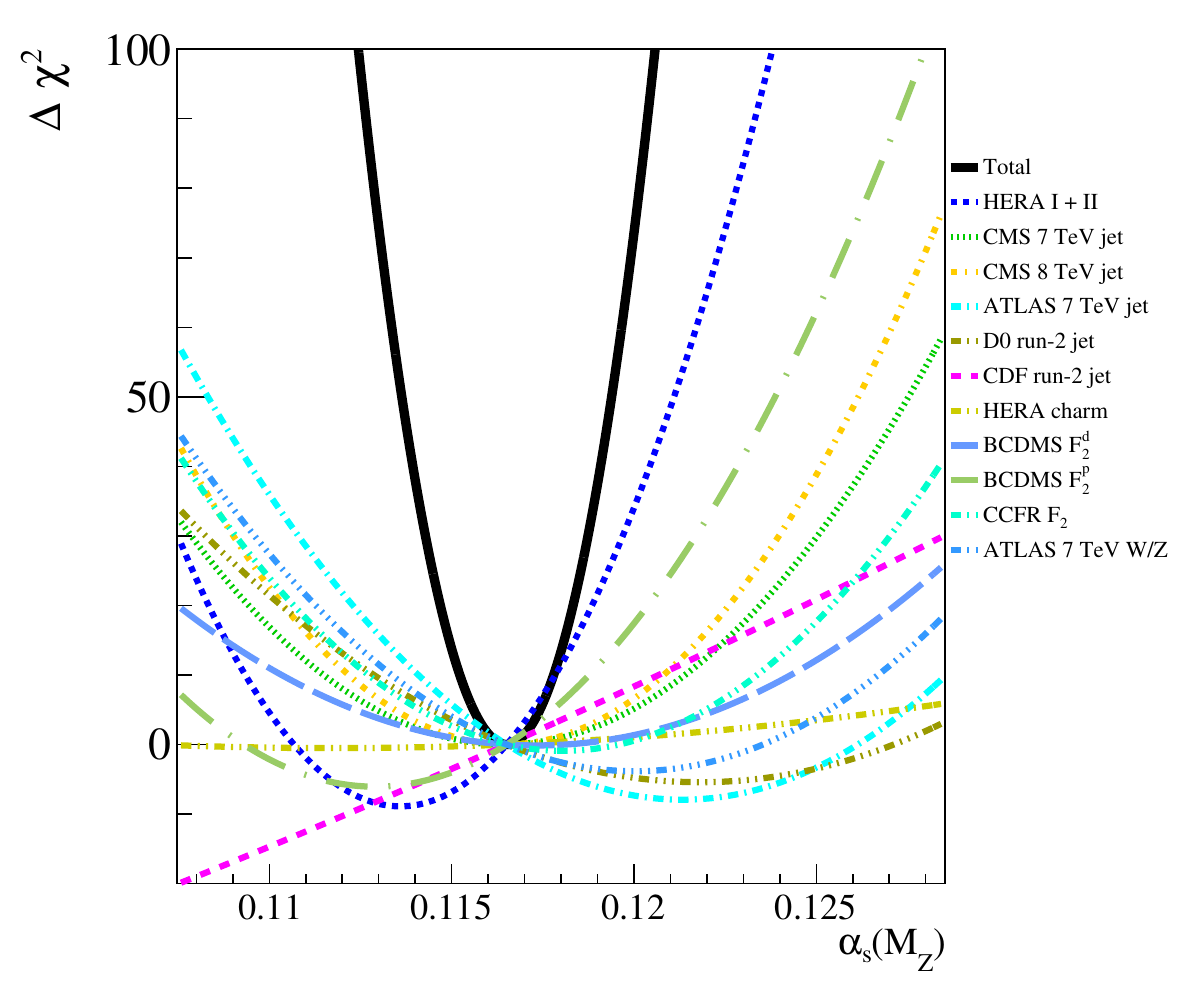}
    \includegraphics[width=0.48\linewidth]{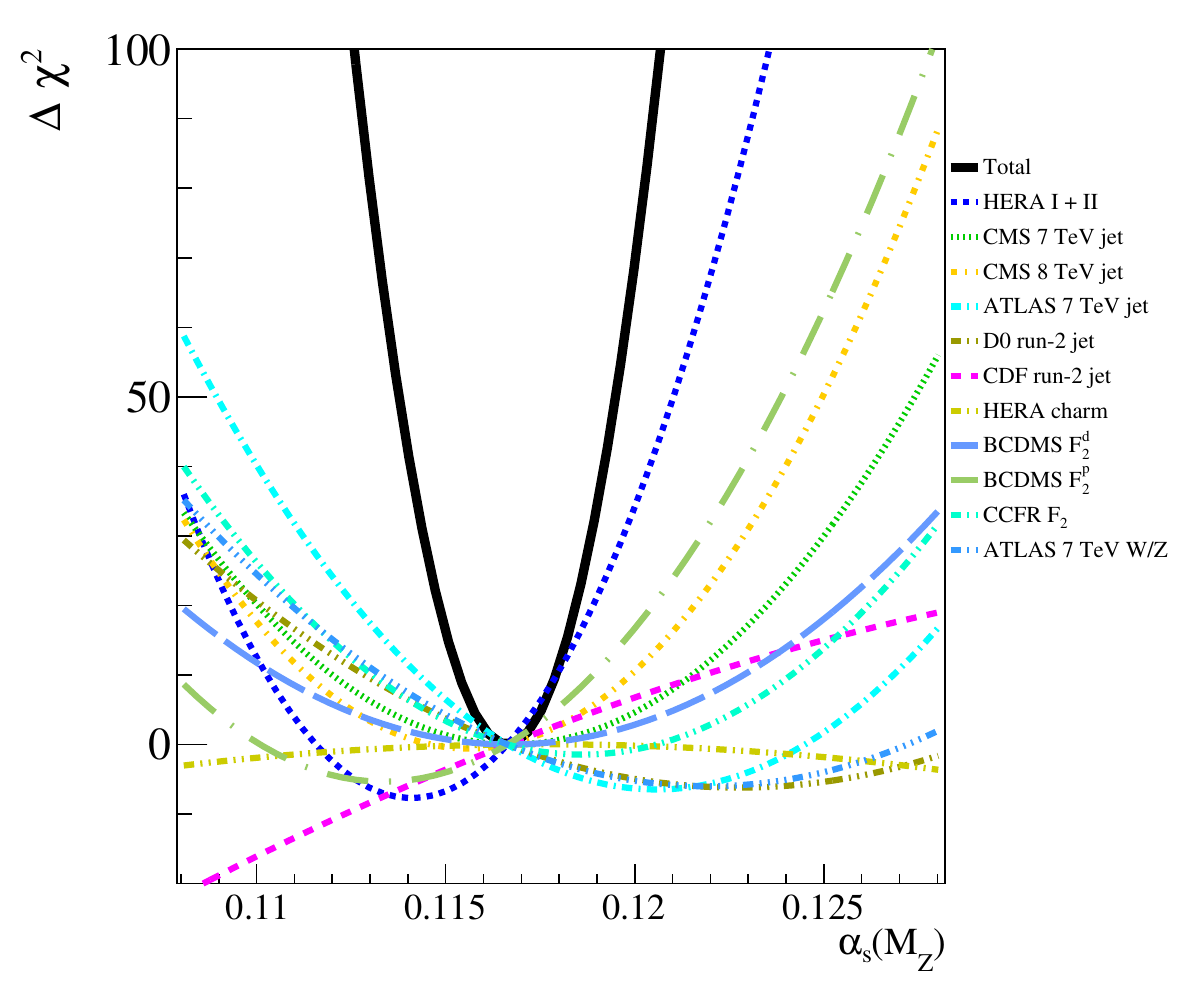}
    \caption{The scan of the strong coupling constant at the scale $M_Z$ for NNLO PDFs. Left: The baseline PDFs are obtained from the standard CT18A analysis but with the E248 data set (ATLAS 7 TeV $W/Z$ production with an integrated luminosity of 4.6~fb$^{-1}$ collected in 2011~\cite{ref:E248}) removed, and subsequently updated by \texttt{ePump} to include the missing E248 data. The $\chi^2$ values are computed using \texttt{ePump}. Right: The corresponding results obtained from the full CT18A analysis, using the CTEQ-TEA  global analysis code.}
    \label{fig:AlphaS_scan_248}
\end{figure}

Figure~\ref{fig:AlphaS_scan_268} compares the $\alpha_s$ scans obtained from \texttt{ePump}, where the CT18Am248 PDFs are updated with the E268 data set, with those from the full CT18 global analysis. (Data sets with negligible sensitivity to $\alpha_s$ are not shown.)
While small quantitative differences are observed for some data sets—reflecting the absence of higher-order nonlinear effects in the \texttt{ePump} linear approximation—the overall trends in the $\alpha_s$  dependence are consistent between the two approaches.
Figure~\ref{fig:AlphaS_scan_248} shows a similar comparison for the case where the PDFs are updated with the E248 data set, and the corresponding full global analysis is the CT18A fit.
The upgraded \texttt{ePump} accurately reproduces the shifts in PDFs, the preferred $\alpha_s(m_Z)$, and the associated uncertainty reductions obtained in full global fits, as confirmed by comparison with Lagrange--Multiplier scans. Small deviations are observed only for data sets whose $\chi^2$ profiles exhibit nonlinear behavior—such as the CMS 8~TeV~\cite{ref:E545} inclusive jet data—where the quadratic Hessian approximation is insufficient to capture the full shape of the $\chi^2$ profile near the best-fit point.\footnote{Nonlinear effects on PDF uncertainties can in principle be estimated systematically~\cite{Zhan:2024tic}, but such a treatment is beyond the scope of the present \texttt{ePump} implementation.}
We note that, within the \texttt{ePump} approximation, the $\Delta\chi^2$ profile obtained from the Lagrange-Multiplier scan is parabolic by construction, whereas the corresponding profile from a full global analysis need not be parabolic when nonlinear effects are significant for a given data set.

\section{Physics Examples}\label{sec:demo}

To illustrate the practical capabilities of the upgraded \texttt{ePump} framework, we present three representative applications drawn from common use cases in collider phenomenology. 
First, we assess the impact of newly available precision measurements, not included in CT18, on the extracted value of $\alpha_s(m_Z)$ and its uncertainty. Second, we compare the gluon PDF obtained from PDF-only updating with that from simultaneous PDF+$\alpha_s$ updating, in order to quantify the effect of treating $\alpha_s$ as a free parameter. Third, we examine the $gg \rightarrow H$ production cross section under both updating schemes and display the resulting correlation ellipse between $\sigma(gg \rightarrow H)$ and $\alpha_s$.

\subsection{Impact of New Data Sets on $\alpha_s$}

After validating the performance of the \texttt{ePump} framework against full global re-analyses, we proceed to study the impact of additional experimental measurements that were not included in the CT18 global analysis. The following post-CT18 data sets, labeled according to the experimental dataset numbering convention of CT25~\cite{ref:CT25AS}, are incorporated using \texttt{ePump} to perform simultaneous updates of the PDFs and the strong coupling constant $\alpha_s$:

\begin{itemize}
	\item E218 (LHCb 13~TeV, with an integrated luminosity of 0.29~fb$^{-1}$; 
	$Z$-boson cross sections with $p_{T}^\ell > 20$~GeV, $2.0 < \eta < 4.5$, 
	and $60 < m_{\ell\ell} < 120$~GeV~\cite{ref:E218}),
	\item E553 (ATLAS 8~TeV, with an integrated luminosity of 20.3~fb$^{-1}$; 
	inclusive jet cross sections,  $\frac{d^2\sigma}{dp_Td|y|}$, 
	$|y| < 3.0$, $p_T^{\text{jet}} \in [70,\,2500]$~GeV, $R = 0.6$~\cite{ref:E553}),
	\item E554 (ATLAS 13~TeV, with an integrated luminosity of 3.2~fb$^{-1}$; 
	inclusive jet cross sections, $\frac{d^2\sigma}{dp_Td|y|}$, 
	$|y| < 3.0$, $p_T^{\text{jet}} \in [100,\,3937]$~GeV, $R = 0.4$~\cite{ref:E554}).
\end{itemize}

After including these measurements, the preferred value of the strong coupling shifts from
\[
\alpha_s(m_Z) = 0.1164 \pm 0.0025
\]
to
\[
\alpha_s(m_Z) = 0.1171 \pm 0.0024,
\]
where the quoted uncertainties correspond to the 68\% confidence level. 
The addition of the new data leads to a moderate upward shift in the central value, accompanied by a slight reduction of the uncertainty.
This behavior reflects the sensitivity of high-precision electroweak boson and jet production data to $\alpha_s$. In particular, the inclusive jet measurements provide direct constraints on the gluon distribution, which is strongly correlated with the determination of the strong coupling in the global fit. 

\begin{figure}[H]
    \centering
    \includegraphics[width=0.55\linewidth]{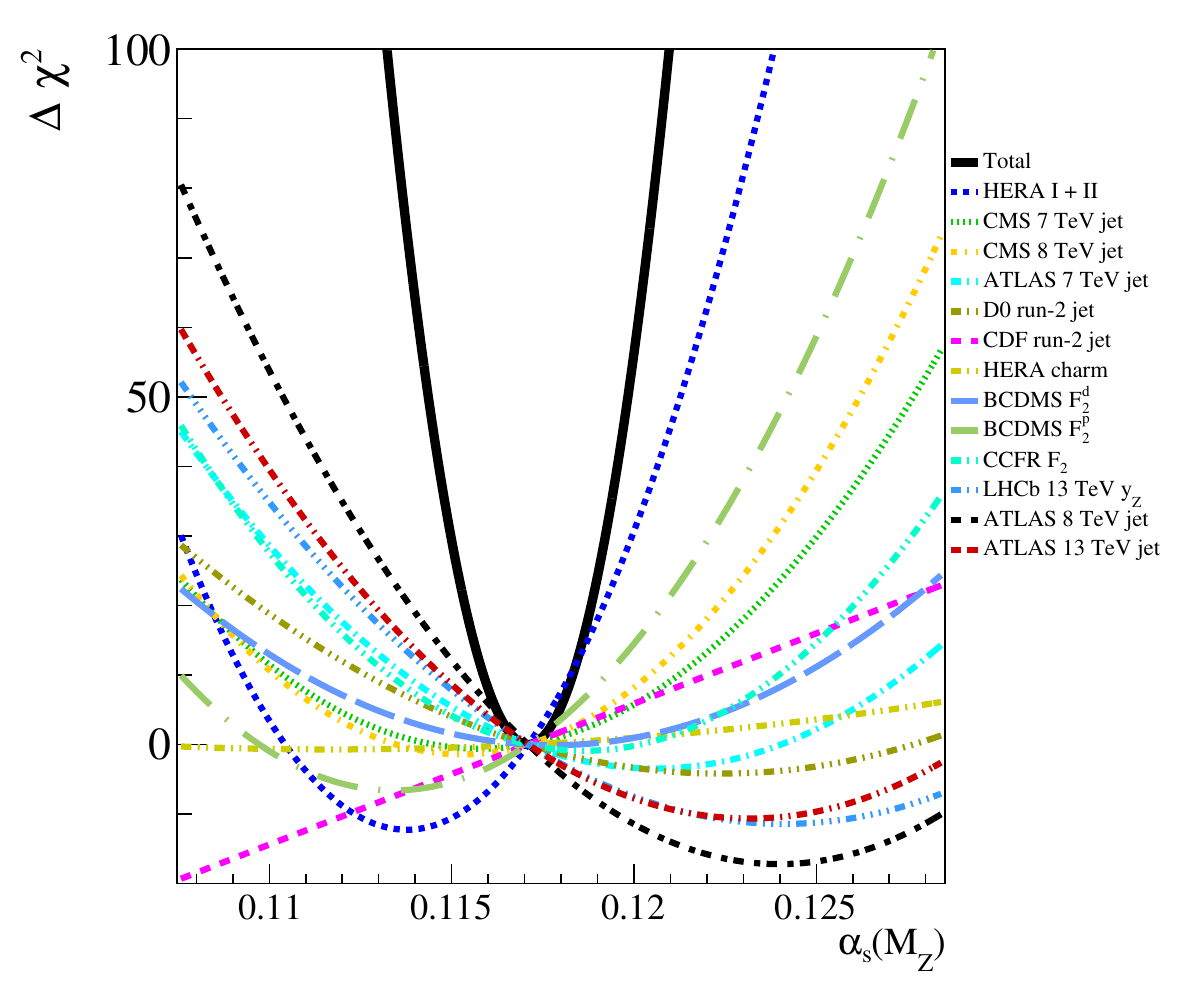}
    \caption{The scan of the strong coupling constant at the scale $M_Z$ for NNLO PDFs. The baseline PDFs are obtained from the standard CT18A analysis but with the E248 data set (ATLAS 7 TeV $W/Z$ production  with an integrated luminosity of 4.6~fb$^{-1}$ collected in 2011) removed, and subsequently updated by \texttt{ePump} to include the new E218, E553, and E554 data sets. The $\chi^2$ values are computed using \texttt{ePump}. The preference for a larger $\alpha_s(m_Z)$ is driven consistently by all three post-CT18 data sets: E218 (LHCb 13 TeV $y_Z$), E553 (ATLAS 8 TeV jet), and E554 (ATLAS 13 TeV jet).} 
    	
    \label{fig:AlphaS_scan_218_553_554}
\end{figure}

To further quantify the constraint imposed by the new data, we perform $\alpha_s$ scans within the \texttt{ePump} framework after the update. Figure~\ref{fig:AlphaS_scan_218_553_554} shows the resulting $\chi^2$ profiles, which exhibit a shift of the minimum toward larger $\alpha_s$ values and a slightly increased curvature relative to the baseline scan, indicating a tightened constraint on $\alpha_s$. The preference for a larger $\alpha_s(m_Z)$ is driven consistently by all three post-CT18 data sets—E218, E553, and E554—and is in agreement with the findings of a recent full global fit performed in the CT25 framework~\cite{ref:CT25AS}.

These studies demonstrate that the extended \texttt{ePump} framework provides a practical and efficient tool for rapidly evaluating the impact of new measurements on the correlated determination of PDFs and $\alpha_s$, without the need for a full global refit.

\subsection{Gluon PDF Sensitivity to Alternative Data Sets}

To further investigate the impact of the correlated PDF+$\alpha_s$ updating on the gluon distribution, we perform an additional study using a different selection of data sets that provide stronger constraints on the gluon PDF.
The selected data sets include measurements with significant sensitivity to the gluon distribution via both scaling violations and gluon-initiated subprocesses. In this study, the data sets incorporated in the \texttt{ePump} update are:

\begin{itemize}
	\item E248 (ATLAS 7~TeV, with an integrated luminosity of 4.6~fb$^{-1}$; 
	combined $W/Z$ cross sections~\cite{ref:E248}),	
	\item E553 (ATLAS 8~TeV, with an integrated luminosity of 20.3~fb$^{-1}$; 
	inclusive jet cross sections, $\frac{d^2\sigma}{dp_Td|y|}$, 
	$|y| < 3.0$, $p_T^{\text{jet}} \in [70,\,2500]$~GeV, $R = 0.6$~\cite{ref:E553}),	
	\item E581 (CMS 13~TeV, with an integrated luminosity of 137~fb$^{-1}$; 
	top‑quark pair production, $m_{t\bar{t}}$ distribution in the lepton+jet channel~\cite{ref:E581}),
	\item E528 (CMS 13~TeV, with an integrated luminosity of 35.9~fb$^{-1}$; 
	top‑quark pair production, $y_{t\bar{t}}$ distribution in the dilepton channel~\cite{ref:E528}).
\end{itemize}

\begin{figure}[H]
    \centering
    \includegraphics[width=0.5\linewidth]{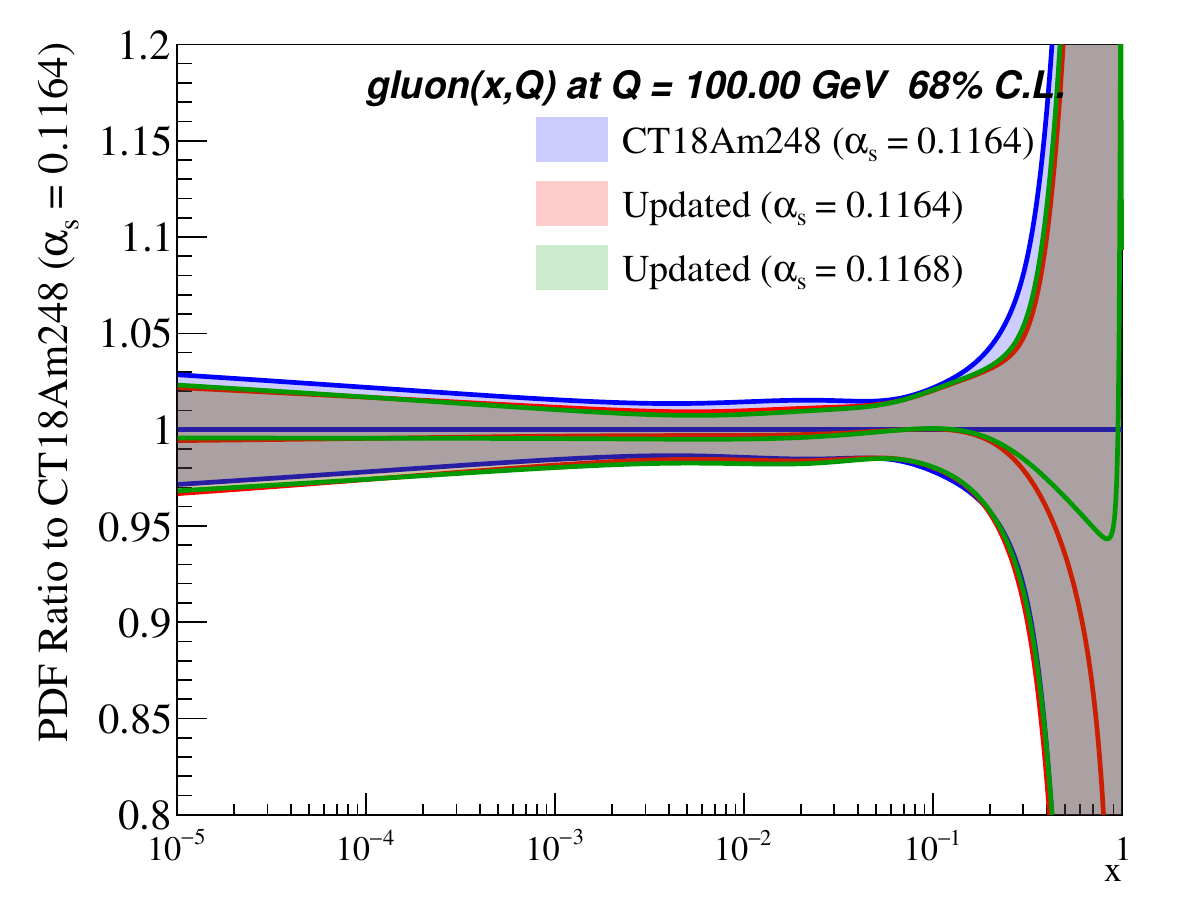}
    \caption{Comparison of the gluon PDF for three analyses. The first (CT18Am248) corresponds to the standard CT18A analysis with the E248 data set removed. The middle result is obtained by updating this baseline using \texttt{ePump} to include E248 and the new E553, E581, and E528 data sets. The bottom result is obtained by updating with the same new data sets, but with $\alpha_s(m_Z)$ simultaneously updated.
    }
    \label{fig:gluonPDF}
\end{figure}

Figure~\ref{fig:gluonPDF} shows the comparison of the gluon PDFs obtained from the baseline CT18Am248 set and the PDFs updated with \texttt{ePump} using (i) PDF-only updating and (ii) simultaneous PDF+$\alpha_s$ updating. The inclusion of these gluon-sensitive measurements leads to visible modifications of the gluon distribution, particularly in the medium- and large-$x$ regions.

\begin{figure}[H]
    \centering
    \includegraphics[width=0.32\linewidth]{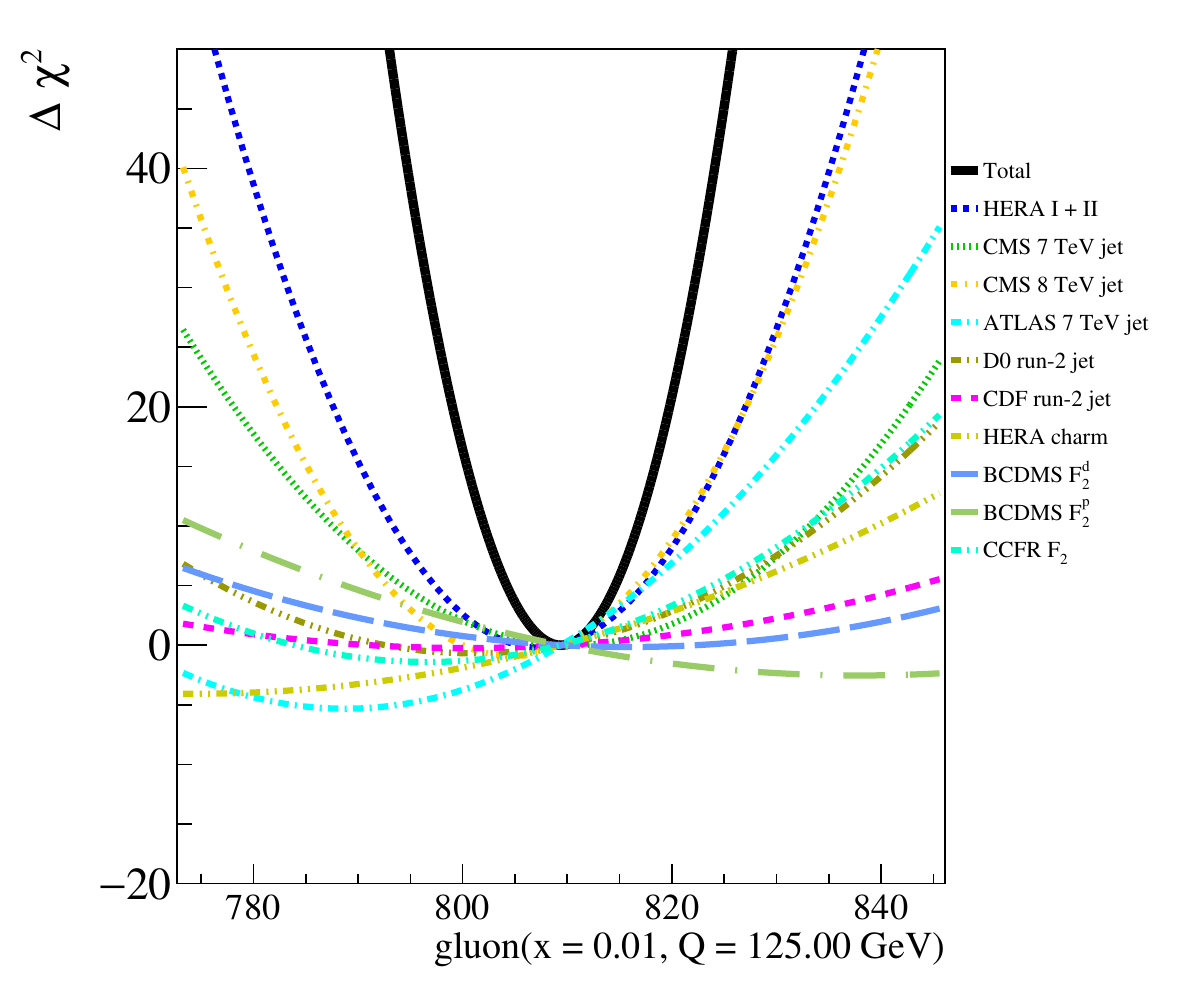}
    \includegraphics[width=0.32\linewidth]{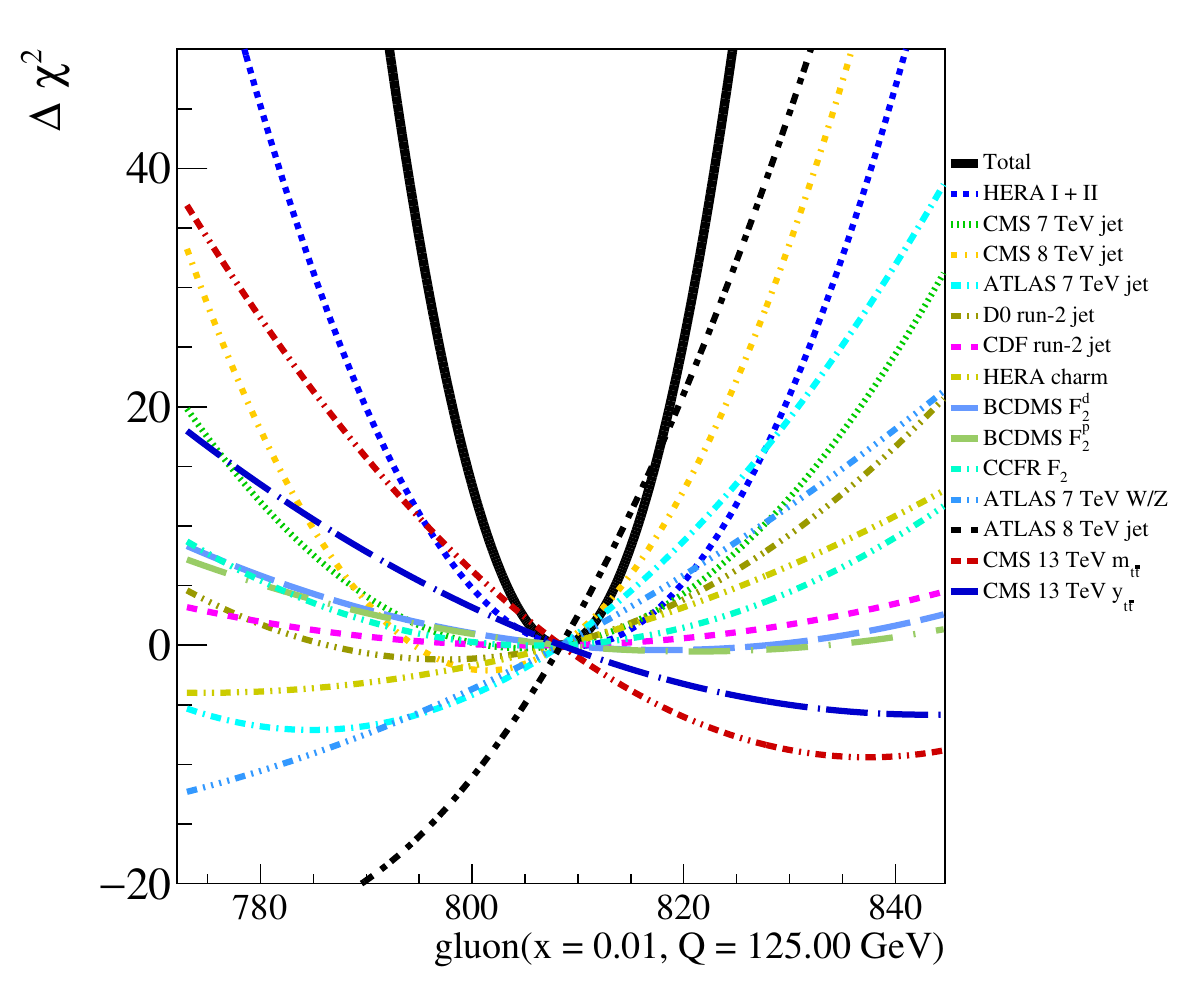}
    \includegraphics[width=0.32\linewidth]{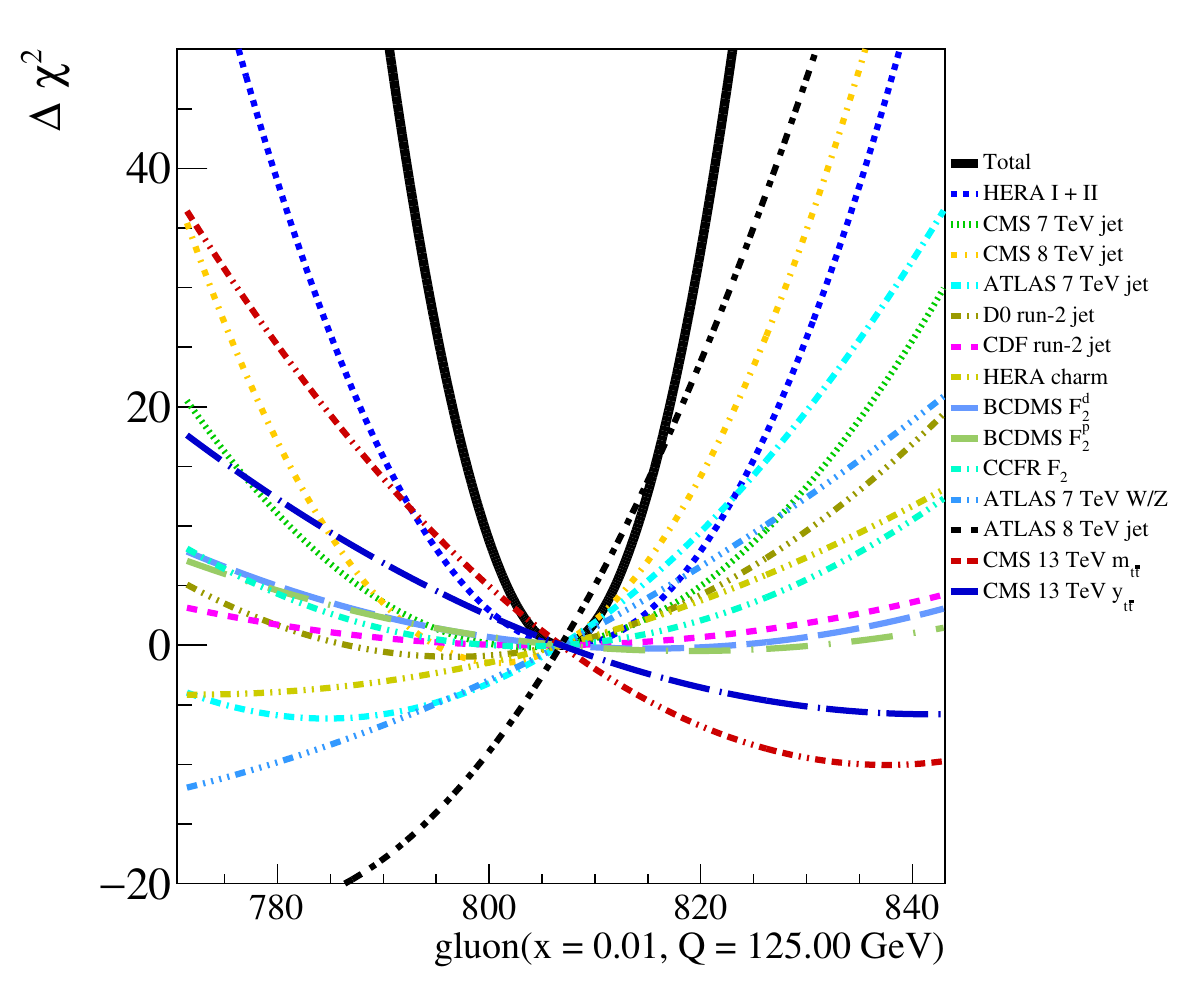}
    \caption{LM scans for the gluon PDF at $Q = 125$~GeV and $x = 0.01$. 
    	Left: The PDF is based on the standard CT18A analysis with the E248 data set removed. 
    	Middle: The PDF is updated by \texttt{ePump} to include E248 (ATLAS 7 TeV $W/Z$) and the new E553 (ATLAS 8 TeV jet), E581 (CMS 13 TeV $m_{t {\bar t}}$), and E528  (CMS 13 TeV $y_{t {\bar t}}$) data sets. 
    	Right: The PDF is updated using the same new data sets, but with $\alpha_s$ updated simultaneously. 
    	All $\chi^2$ values are computed using \texttt{ePump}. The CMS 13~TeV top-quark pair production data sets, E581 and E528, consistently prefer an enhanced gluon PDF at small $x$ ($x=0.01$).}
    \label{fig:gluon_scan_0.01}
\end{figure}

\begin{figure}[H]
    \centering
    \includegraphics[width=0.32\linewidth]{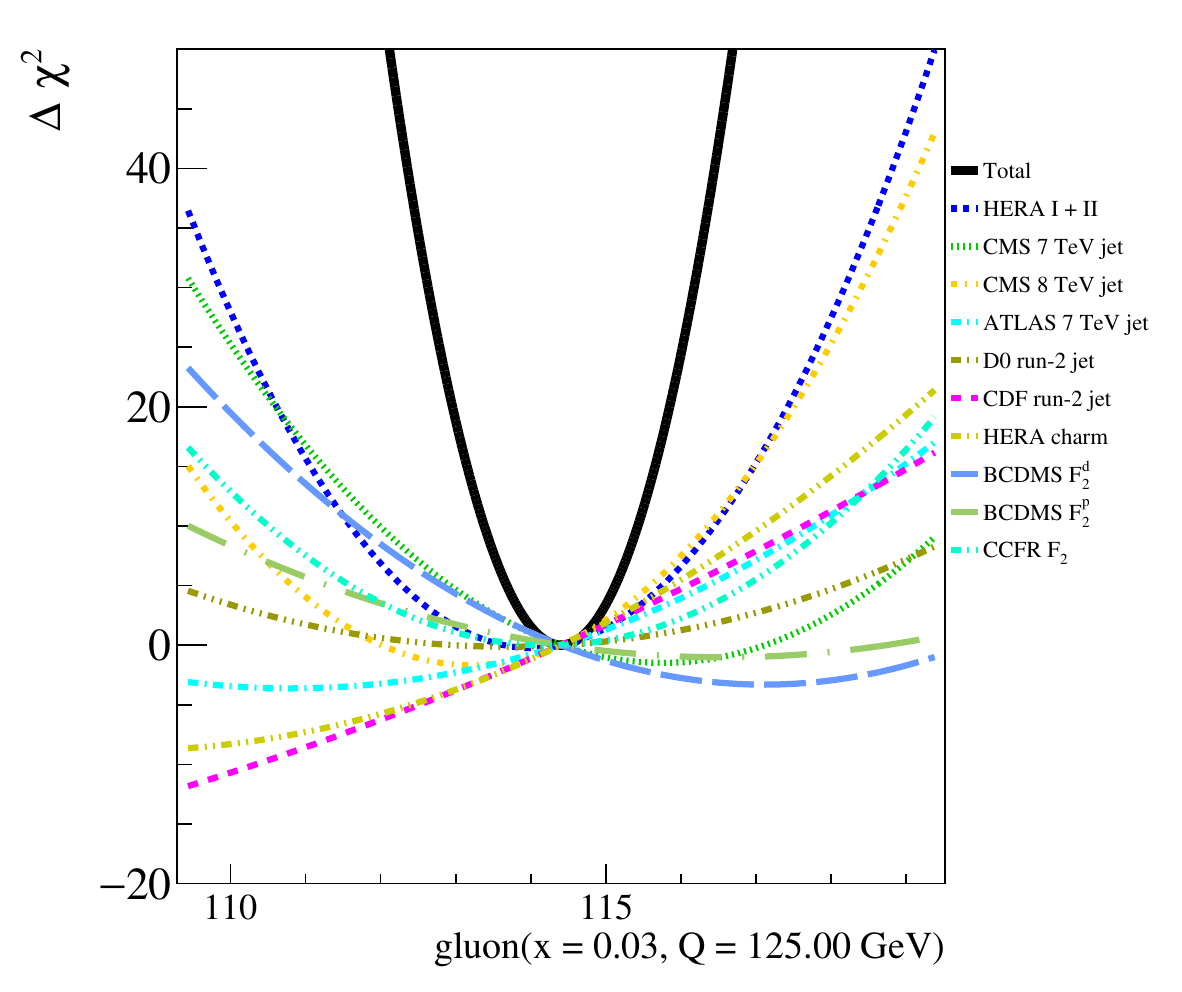}
    \includegraphics[width=0.32\linewidth]{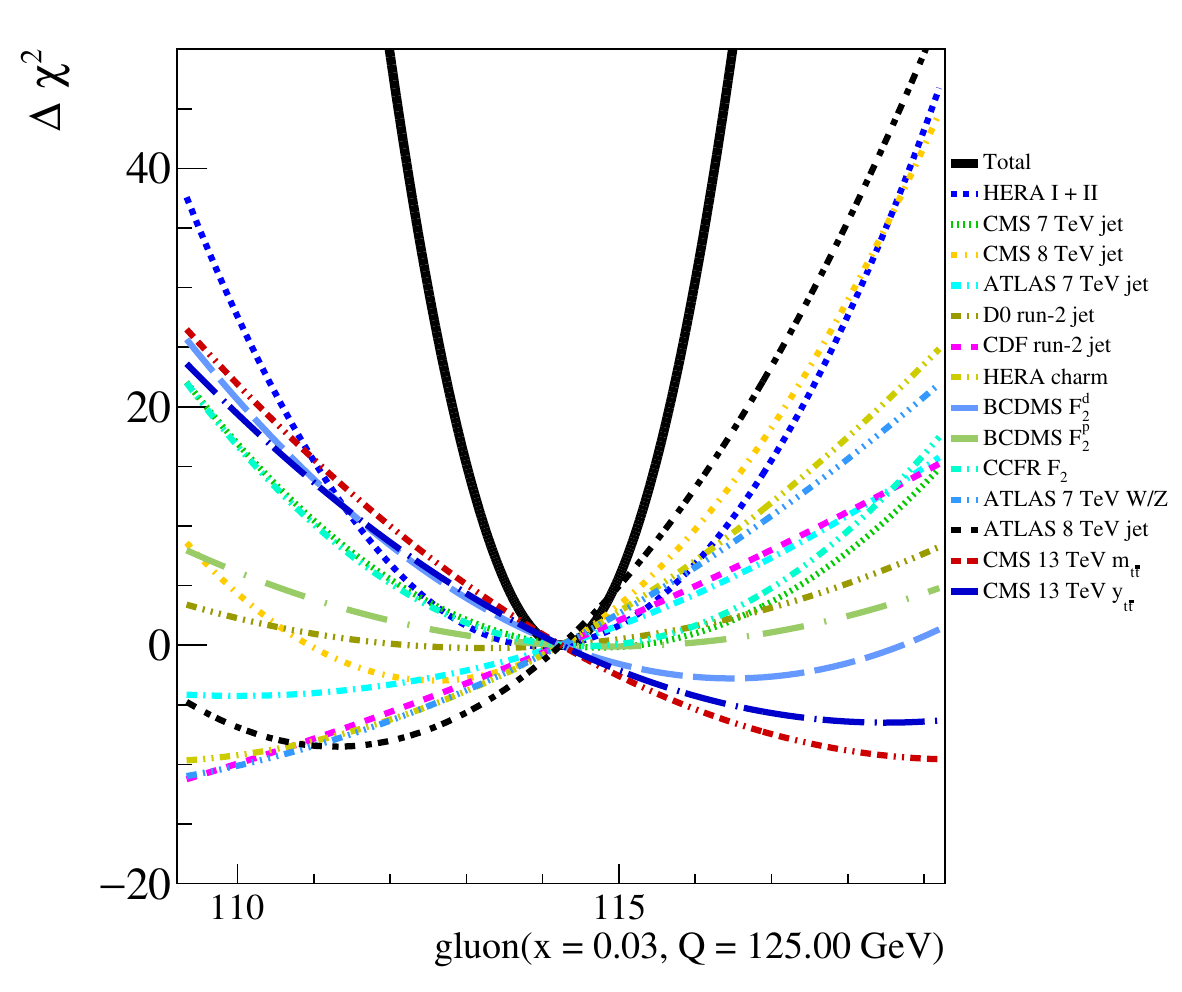}
    \includegraphics[width=0.32\linewidth]{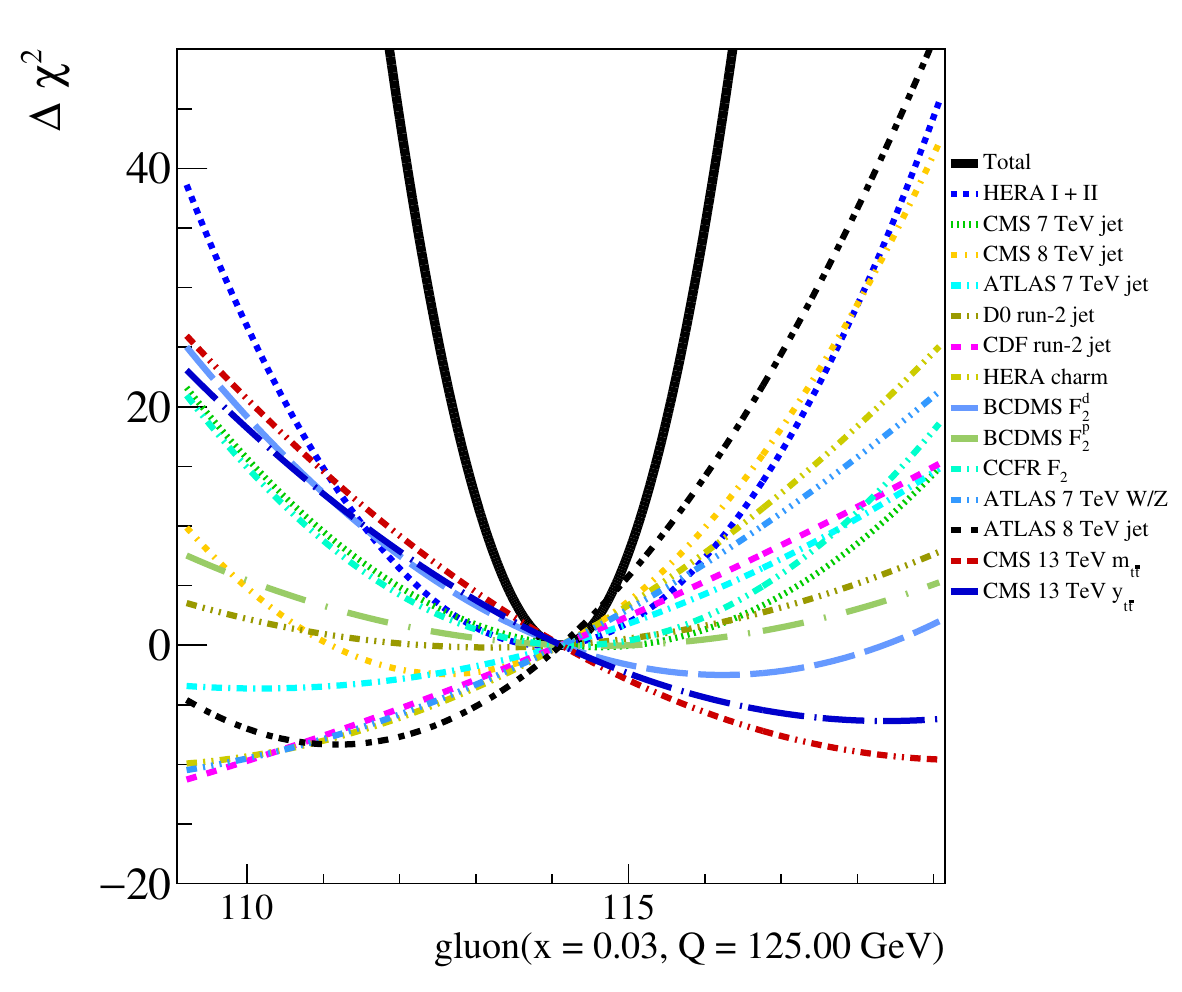}
    \caption{Similar to Fig.~\ref{fig:gluon_scan_0.01}, but for the gluon PDF at $Q=125$~GeV and $x=0.03$.}
    \label{fig:gluon_scan_0.03}
\end{figure}

\begin{figure}[H]
    \centering
    \includegraphics[width=0.32\linewidth]{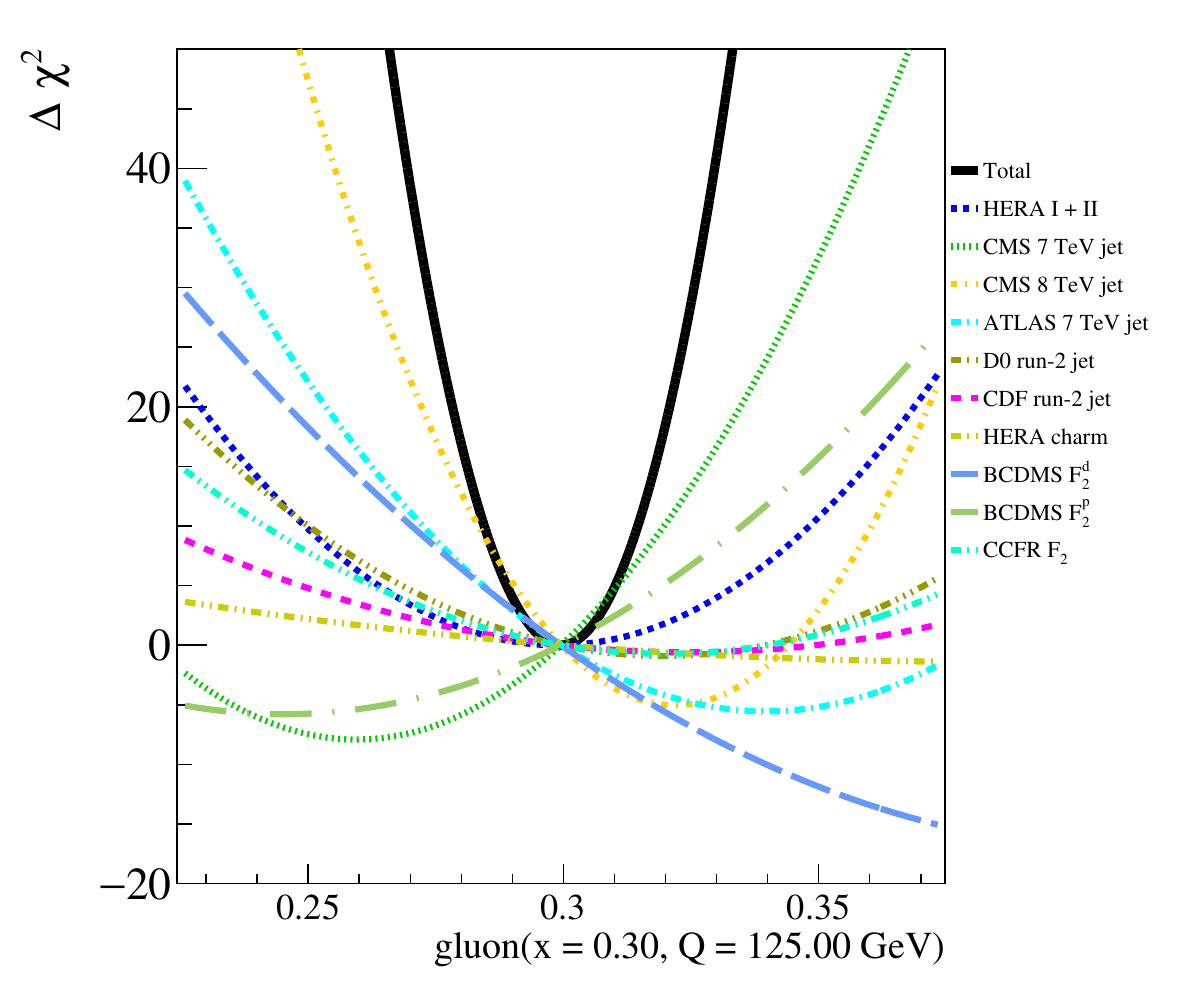}
    \includegraphics[width=0.32\linewidth]{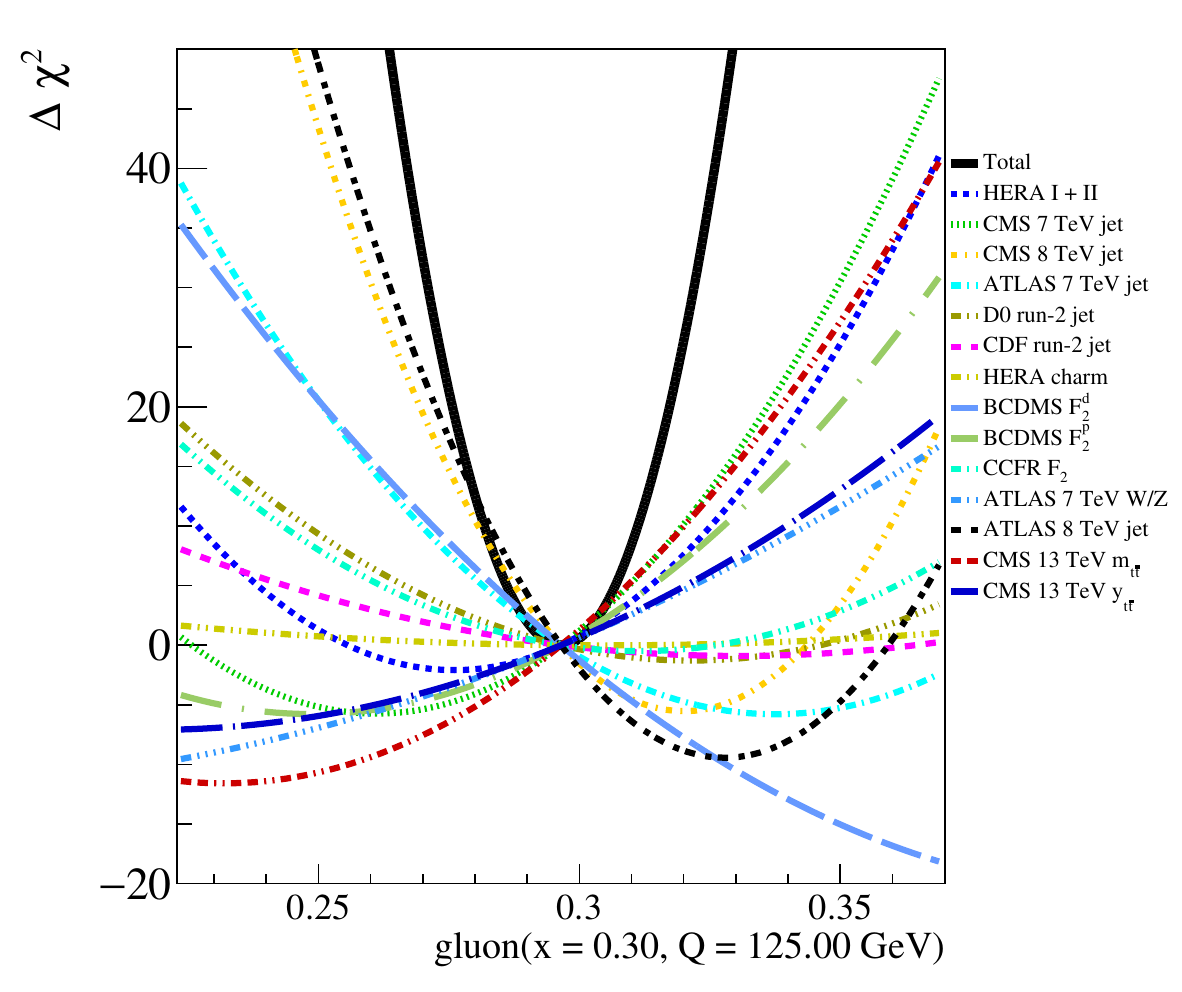}
    \includegraphics[width=0.32\linewidth]{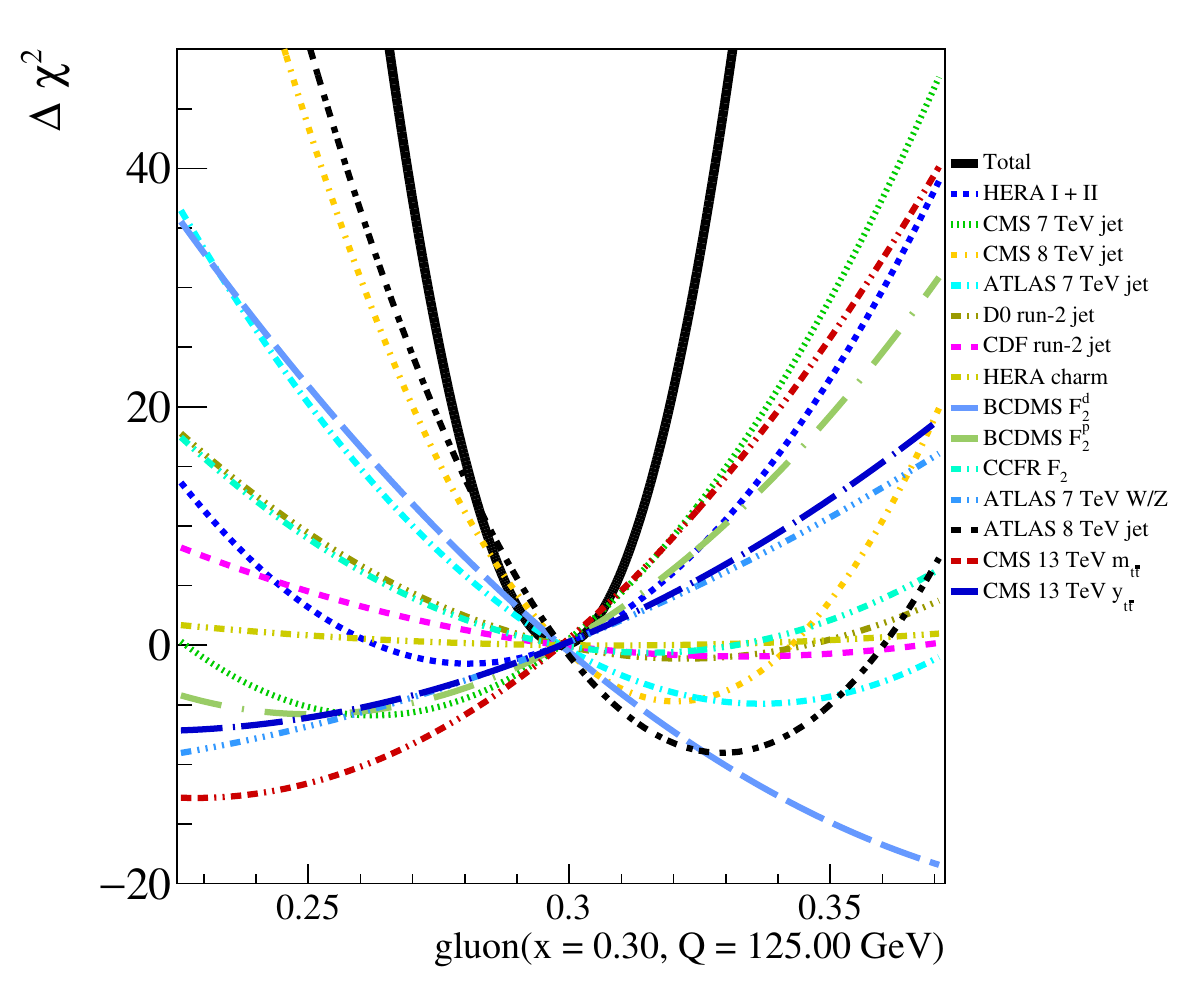}
    \caption{Similar to Fig.~\ref{fig:gluon_scan_0.01}, but for the gluon PDF at $Q=125$~GeV and $x=0.3$. The CMS 13~TeV top-quark pair production data sets, E581 and E528, consistently prefer a suppressed gluon PDF at large $x$ ($x=0.3$).}
    \label{fig:gluon_scan_0.3}
\end{figure}

To quantify this behavior, Lagrange-Multiplier scans of the gluon PDF are performed at representative values of $x$ and $Q$. Figures~\ref{fig:gluon_scan_0.01}, \ref{fig:gluon_scan_0.03}, and \ref{fig:gluon_scan_0.3} show the $\chi^2$ profiles of the gluon distribution at $x=0.01$, $0.03$, and $0.3$, respectively, for the baseline CT18Am248 PDF set, the PDF-only update, and the simultaneous PDF+$\alpha_s$ update. The CMS 13~TeV top-quark pair production data sets, E581 and E528, consistently prefer an enhanced gluon PDF at small $x$ ($x=0.01$ and $0.03$) and a suppressed gluon PDF at large $x$ ($x=0.3$), evaluated at $Q=125$~GeV.

These results demonstrate that the impact of the correlated PDF+$\alpha_s$ updating becomes increasingly important when the added measurements provide strong constraints on the gluon distribution. 
This underscores the importance of treating $\alpha_s$ and the PDFs consistently when incorporating gluon-sensitive data.

\subsection{$\sigma(gg\rightarrow H)$}

Beyond the direct examination of the gluon PDF, it is useful to study the impact of the \texttt{ePump} updating on a physical observable that is simultaneously sensitive to both the gluon distribution and the strong coupling constant $\alpha_s$.
The Higgs boson production cross section via gluon fusion, $\sigma(gg\rightarrow H)$, provides an ideal probe for this purpose.
In this subsection, we use the same data combination as in the preceding section: starting from the CT18Am248 baseline (obtained by removing E248 from the CT18A analysis), we apply \texttt{ePump} updates including the data sets E248, E553, E581, and E528, under both the PDF-only and simultaneous PDF+$\alpha_s$ scenarios.
The $gg\rightarrow H$ cross section is calculated at N$^3$LO accuracy in QCD using the \texttt{ggHiggs} package~\cite{ref:ggHiggs1,ref:ggHiggs2,ref:ggHiggs3,ref:ggHiggs4}, with the factorization and renormalization scales both set equal to the Higgs boson mass, $m_H = 125$~GeV. Table~\ref{tab:ggHiggs} summarizes the predicted cross sections and the corresponding values of $\alpha_s$ before and after the \texttt{ePump} updates. The second and third rows show the results for the PDF-only updating and the simultaneous PDF+$\alpha_s$ updating, respectively.
\begin{table*}[hbt]
    \centering
    \begin{tabular}{c|c|c}
        \hline \hline
        &  $\sigma(gg \rightarrow H)$~[pb] & $\alpha_s$ \\
        \hline
        Before update & 40.21 $\pm$ 0.71 (PDF) $\pm$ 1.91 ($\alpha_s$) & 0.1164 $\pm$ 0.0025 \\
        After update (PDF only) & 39.98 $\pm$ 0.64 (PDF) $\pm$ 1.91 ($\alpha_s$) & 0.1164 $\pm$ 0.0025 \\
        After update (PDF + $\alpha_s$) & 40.28 $\pm$ 0.64 (PDF) $\pm$ 1.78 ($\alpha_s$) & 0.1168 $\pm$ 0.0023 \\
        \hline \hline
    \end{tabular}
    \caption{Predictions for the Higgs boson ($m_H = 125$~GeV) production cross section via gluon fusion, $\sigma(gg \rightarrow H)$, together with the corresponding $\alpha_s(m_Z)$ values before and after the \texttt{ePump} update, which includes E248 and the new E553, E581, and E528 data sets. The quoted uncertainties correspond to the 68\% CL.}
    \label{tab:ggHiggs}
\end{table*}

Compared with the PDF-only case, the simultaneous updating leads to additional shifts in the predicted cross section, reflecting the correlated changes in the gluon normalization and the value of $\alpha_s$.

\begin{figure}[H]   
 \centering
    \includegraphics[width=0.55\linewidth]{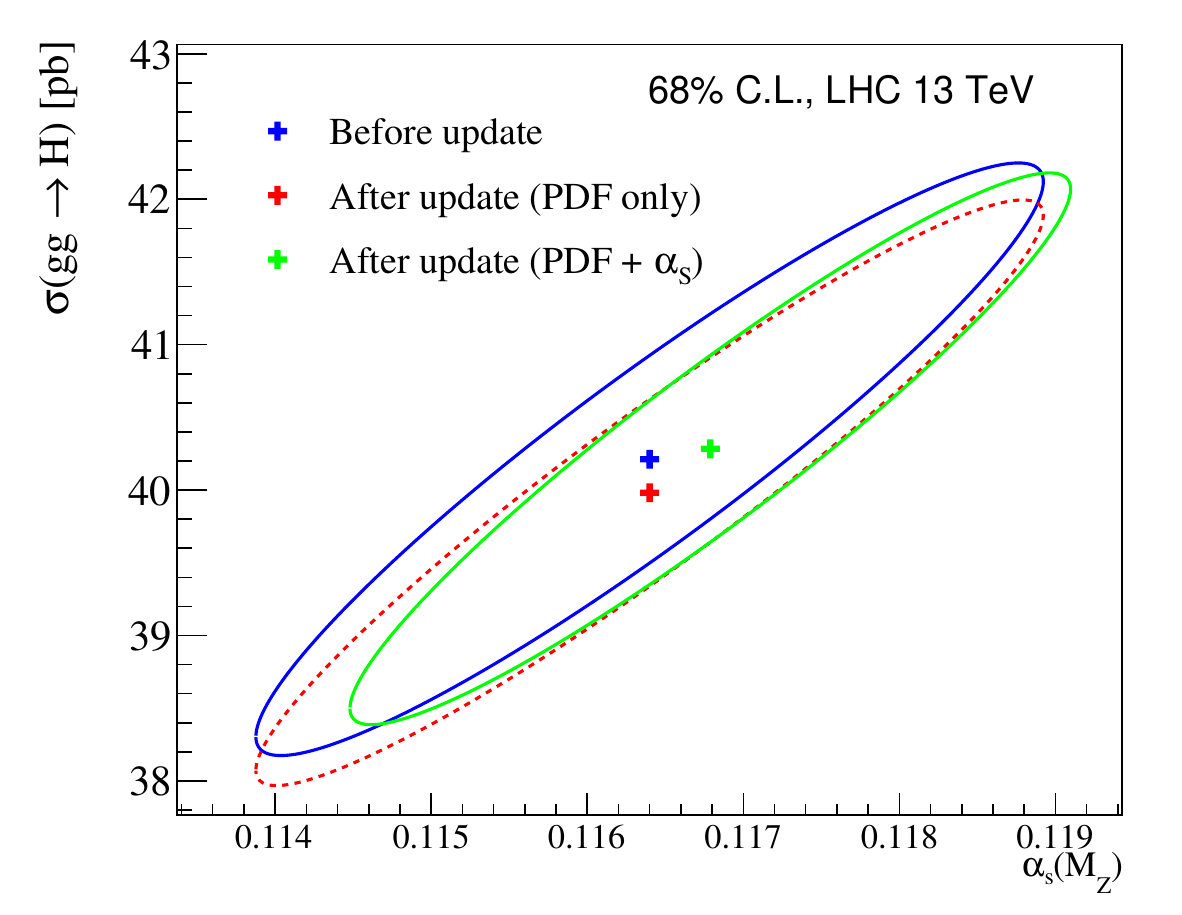}
    \caption{Correlation ellipse between the inclusive Higgs boson production cross section, $\sigma(gg \rightarrow H)$, at the 13~TeV LHC and $\alpha_s(m_Z)$, shown at the 68\% CL. “Before update” denotes the baseline CT18Am248 PDF set, obtained by removing the E248 data from the CT18A analysis. “After update (PDF+$\alpha_s$)” corresponds to the PDF set obtained with \texttt{ePump} after including the E248, E553, E581, and E528 data sets. 
    For comparison, the dashed ellipse labeled “After update (PDF only)” shows the case in which only the PDFs are updated while $\alpha_s(m_Z)$ is fixed at $0.1164$. Consequently, the two $\alpha_s$ eigenvector sets are not updated by the new data.}
    \label{fig:ggHiggsScan}
\end{figure}

Figure~\ref{fig:ggHiggsScan} displays the correlation ellipses for the joint $\{\sigma(gg\rightarrow H),\,\alpha_s\}$ uncertainty at the 13~TeV LHC, before and after the \texttt{ePump} update. The orientation and displacement of the ellipses illustrate the strong positive correlation between $\sigma(gg\rightarrow H)$ and $\alpha_s$, while their shift demonstrates that the simultaneous PDF+$\alpha_s$ update yields a larger preferred value of $\alpha_s(m_Z)$ and reduces both the PDF and $\alpha_s$ uncertainties on $\sigma(gg\rightarrow H)$. These results confirm that correlated PDF+$\alpha_s$ updating can have a non-negligible impact on precision predictions for gluon-dominated processes.
This construction is closely related to the correlation analysis performed in previous CTEQ studies~\cite{ref:CTHiggsPaper}, where contours of constant $\Delta\chi^2$ in the $(\alpha_s,\sigma_H)$ plane were used to determine the correlated PDF+$\alpha_s$ uncertainties of the Higgs production cross section.
In the Hessian approximation adopted here, the ellipse corresponds to the quadratic expansion of the global $\chi^2$ around its minimum, such that the contour represents the region of constant $\Delta\chi^2$ for the two-dimensional parameter space spanned by $\alpha_s$ and the observable. The orientation of the ellipse reflects the correlation between $\alpha_s$ and the gluon-fusion cross section, while the extrema along the contour determine the combined PDF+$\alpha_s$ uncertainty of the prediction.
Thus, the present treatment provides a linearized implementation of the same underlying statistical framework used in earlier global analyses. A detailed description of the construction of the observable–$\alpha_s$ correlation ellipse within the \texttt{ePump} framework is given in Appendix~\ref{sec:Ellipse}.

The examples presented in this section demonstrate that the extended \texttt{ePump} framework provides a powerful and computationally lightweight alternative to full global fitting for evaluating the impact of new data on both $\alpha_s$ and the proton PDFs. The method is fast, numerically stable, internally consistent within the Hessian framework, and reliably captures the essential physics encoded in high-precision collider measurements.

\section{Conclusion}
\label{sec:conclusion} 

We have presented and validated a significant extension of the \texttt{ePump}
Hessian-profiling framework that enables the \emph{simultaneous} updating of
proton PDFs and the strong coupling constant $\alpha_s$ in response to new
experimental data. 
By promoting $\alpha_s$ to a dynamical parameter within the
profiling formalism, the method consistently determines correlated shifts in
$\{\text{PDFs}, \alpha_s\}$ while preserving the full covariance structure of the original global analysis. 
In practice, this treats $\alpha_s$ on the same
footing as the PDF parameters, incorporating its effects both in DGLAP
evolution and in perturbative cross-section predictions.

The resulting framework provides a computationally efficient approximation to
a full global QCD refit in the combined $\{\text{PDFs}, \alpha_s\}$  parameter space. Through a coordinated
set of validation studies—including comparisons with full global re-analyses,
Lagrange–Multiplier scans, and detailed uncertainty-propagation checks—we find
that the extended \texttt{ePump} reliably reproduces the joint behavior of
$\{\text{PDFs}, \alpha_s\}$ that would otherwise require substantially greater
computational resources. In particular, it captures the interplay between
$\alpha_s$ and the gluon distribution and preserves the structure of correlated
uncertainties across diverse collider and DIS inputs.

We illustrated the phenomenological utility of the method using representative collider measurements that simultaneously constrain both the gluon PDF and $\alpha_s$. These
studies show that correlated updates of PDFs and $\alpha_s$ can produce 
measurable shifts in precision observables, as exemplified by the Higgs boson
production cross section via gluon fusion. As experimental precision at the
LHC continues to improve—and with the much larger datasets anticipated from
the High-Luminosity LHC—the ability to assess the impact of new data rapidly
and consistently will become increasingly essential.

The extended \texttt{ePump} framework thus fills an important role in the
precision-frontier era: it provides a fast, practical, and robust tool for
quantifying the physics impact of new measurements, informing experimental
design, guiding theory interpretation, and shaping global-fit strategy,  without the overhead of a full refit. 
Looking ahead, natural directions include extensions to mixed QCD–electroweak effects, systematic studies of higher-order scale variations, and combined PDF–SMEFT profiling, where SMEFT denotes the Standard Model Effective Field Theory framework in which potential new-physics effects are parametrised by higher-dimensional operators. Together, these developments will support a more agile and data-responsive global analysis program, advancing quantitative tests of QCD and enabling the highest-precision probes of the Standard Model.

\acknowledgments
We thank Alim Ablat, Sayipjamal Dulat, Tie-Jiun Hou, and Keping Xie for their contributions during the early stages of this work, and our CTEQ-TEA colleagues for many helpful discussions. This work was supported by the U.S. National Science Foundation under Grant No. PHY-2310291.

\clearpage

\appendix

\section{Dynamical Tolerances in the PDF$+\alpha_s$ Analysis with \texttt{ePump}}\label{sec:Dynamical Tolerances}

We now show that the modifications arising from dynamical tolerances in the fixed-$\alpha_s$ PDF updating carry through directly to the full PDF$+\alpha_s$ updating analysis.
The standard Hessian PDFs now correspond to evaluating functions of the parameters at ${\bf z}^{\pm j}=\pm(T_j^\pm/T){\bf e}^j$ and  $z_\alpha=0$, which gives corresponding values of $\chi^2=(T_j^\pm)^2$. For the moment, assuming that $\alpha_s^0=\alpha_\mathrm{GA}$ and $\delta\alpha_s=\delta\alpha_\mathrm{GA}$, then the $\pm\alpha_s$ PDFs are evaluated by minimizing $\chi^2$ as a function of $z_i$ for fixed $z_\alpha=\pm1$ or $\alpha_s=\alpha_\mathrm{GA}\pm\delta\alpha_\mathrm{GA}$.  Assuming that the $\chi^2$ is exactly quadratic in $\alpha_s$, it will have a value of
$(T^\pm_\alpha)^2\equiv (T_\alpha)^2$ for these choices (relative to its value at $z_\alpha=0$).  Thus, we can write this as 
\begin{eqnarray}
\Delta\chi^2({\bf z},z_{\alpha})&=&T^2\sum_{i=1}^N(z_i-b_iz_\alpha)^2+T_\alpha^2z_\alpha^2\,.\label{eq:chisquarebar}
\end{eqnarray}
Note that the curvature in $\alpha_s$, in this expression, depends only on the ratio $(T_\alpha/\delta\alpha_\mathrm{GA})$.  
Thus, the choice of dynamical tolerance $T_\alpha$ is directly correlated with the extracted uncertainty $\delta\alpha_\mathrm{GA}$. We make the treatment of $\alpha_s$ formally the same as the PDF variables, in the presence of the dynamical tolerances, by defining

\begin{eqnarray}
{z}_{N+1}^\prime&=&\left(\frac{T_\alpha}{T}\right)z_\alpha\nonumber\\
{z}^\prime_i&=&z_i-b_iz_\alpha\nonumber\,,
\label{eq:barpars_zi}
\end{eqnarray}
so that
\begin{eqnarray}
\Delta\chi^2({\bf z},z_{\alpha})&=&T^2\sum_{i=1}^{N+1}(z^\prime_i)^2\,.\label{eq:chisquarebarTiii}
\end{eqnarray}
Note that in the end, nothing should depend on the variable $T^2$,  since it will scale out of every measurable quantity and will be effectively replaced by the dynamical tolerances.

Following the general procedure developed in the original \texttt{ePump} paper, we then obtain, for a general observable, 
\begin{eqnarray}
X({\bf z},z_\alpha)&=&X\bigl({\bf z}^\prime+{\bf b}z_{N+1}^\prime(T/T_\alpha),z_{N+1}^\prime(T/T_\alpha)\bigr)
\nonumber\\
&=&X(\mathrm{\bf 0},0)+\sum_{j=1}^{N+1}\widehat{\Delta X}^jz^\prime_j+\cdots\,
 ,\label{eq:observabledt}
\end{eqnarray}
where
\begin{eqnarray}
\widehat{\Delta X}^{j}&=&
\frac{T_j^-}{T_j^++T_j^-}\left(\frac{X_\alpha(f^{+j},\alpha_s^0)-X_\alpha(f^0,\alpha_s^0)}{T_j^+/T}\right)\nonumber\\
&&+\frac{T_j^+}{T_j^++T_j^-}\left(\frac{X_\alpha(f^0,\alpha_s^0)-X_\alpha(f^{-j},\alpha_s^0)}{T_j^-/T}\right)
\,,\label{eq:dxalphaDyn}
\end{eqnarray}
for $i=1,\dots,N$ and
\begin{eqnarray}
\widehat{\Delta X}^{N+1}&=&
\frac{X(f^{+\alpha},\alpha_s^0+\delta\alpha_s)-X(f^{-\alpha},\alpha_s^0-\delta\alpha_s)}{2(T_\alpha/T)}\,.\label{eq:dxalpha_dyn}
\end{eqnarray}
Note that this last formula is consistent with the previous if we assign $T^\pm_{N+1}=T_\alpha$.

At this point we just redo the analysis from the main text, replacing $\Delta X^{j}\rightarrow\widehat{\Delta X}^{j}$  and $\Delta f^{j}\rightarrow\widehat{\Delta f}^{j}$ everywhere.  The formula for $f^0(x,Q_0)_\mathrm{new}$ and $\delta\alpha_{s,\mathrm{new}}$ are  identical to that in the main text. 
Taking into account that ${z}_{N+1}^\prime=({T_\alpha/T})z_\alpha$, we obtain the modified formulae:
\begin{eqnarray}
\alpha^0_{s,\mathrm{new}}&=&\alpha_s^0+({T/T_\alpha})\bar{z}^{\,\prime}_{N+1}\,\delta\alpha_s \, ,\label{eq:bestfitalphadt}
\end{eqnarray}
and
\begin{eqnarray}
f^{\pm\alpha}(x,Q_0)_\mathrm{new}&=&f^0(x,Q_0)_{\rm new}\pm\left(\frac{\delta\alpha_{s,\mathrm{new}}}{\delta\alpha_{s}}\right)\left(\frac{T_\alpha}{T}\right)\Biggl[\widehat{\Delta{f}}^\alpha(x,Q_0)\nonumber\\
&&-\sum_{i,j=1}^NM^{N+1,i}(\widehat{\delta+M})^{-1}_{ij}\,\widehat{\Delta{f}}^j(x,Q_0)\Biggr]\, .\label{eq:newev_dyn}
\end{eqnarray}
As for the updated Hessian error PDFs at $\alpha_s=\alpha_{s\mathrm{new}}$, we follow the prescription given in the original \texttt{ePump} paper~\cite{ref:ePump} to include the effect of the dynamical tolerances.  Note again that the parameter $T$ cancels from all final, physically observable, results.

\section{Relation between the Global-Analysis and World-Average values of $\alpha_s^0$ and $\delta\alpha_s$}
\label{sec:WorldAverage}

In general, the world-average best-fit value and uncertainty of $\alpha_s$ ($\alpha_\mathrm{WA}$ and $\delta\alpha_\mathrm{WA}$) need not coincide with the values ($\alpha_\mathrm{GA}$ and $\delta\alpha_\mathrm{GA}$) obtained from the data included in the PDF global fit alone.
Also, in practice the best-fit PDF set $f^0(x,Q_0)$ and the Hessian PDFs $f^{\pm j}(x,Q_0)$ are usually obtained from a global analysis at the fixed value of $\alpha_s=\alpha_\mathrm{WA}$, with the two additional $\alpha_s$ PDF sets 
$f^{\pm \alpha}(x,Q_0)$ obtained at the values of $\alpha_s=\alpha_\mathrm{WA}\pm\delta\alpha_\mathrm{WA}$.  This can be taken into account in the global analysis by adding a penalty term to the $\chi^2$ function, which takes into account all of the data used to constrain the world-average fitting of $\alpha_s$ that has not already been included in the global analysis.  Assuming that this additional data is uncorrelated with PDF parameters $z^i$, we can write
\begin{eqnarray}
\Delta\chi^2({\bf z},z_{\alpha},k)&=&T^2\sum_{i=1}^N(z_i-b_iz_\alpha)^2+T_\alpha^2\left[z_\alpha^2+k\left(\frac{\alpha_s-\bar\alpha}{\delta\bar\alpha}\right)^2\right]\,.\label{eq:chisquarebarii}
\end{eqnarray}
where we have chosen
\begin{eqnarray}
z_{\alpha}&=&\frac{\alpha_s-\alpha_\mathrm{GA}}{\delta\alpha_\mathrm{GA}}\,,\label{eq:zalphai}
\end{eqnarray}
and we have allowed for dynamical tolerances.
Thus, for $k=0$, this is identical to Eq.~(\ref{eq:chisquarebar}) from Appendix~\ref{sec:Dynamical Tolerances} with $\alpha_s^0=\alpha_\mathrm{GA}$ and  $\delta\alpha_s=\delta\alpha_\mathrm{GA}$.  Note that the tolerance $T_\alpha$ is that used to determine $\delta\alpha_\mathrm{GA}$, the uncertainty on $\alpha_s$ from the global analysis only.  
(Any tolerance assumed for the non-global-analysis data can be absorbed into the ratio $T_\alpha/\delta\bar{\alpha}$ in the above formula.)

By setting $k=1$ we now include the additional data on the world-average fit of $\alpha_s$. 
The parameters $\bar\alpha$ and $\delta\bar\alpha$ are chosen so that the $\chi^2$ function is minimized at $\alpha_s=\alpha_\mathrm{WA}$ with uncertainty $\delta\alpha_\mathrm{WA}$ at the prescribed confidence level, yielding
\begin{eqnarray}
\frac{1}{(\delta\alpha_\mathrm{WA})^2}&=&\frac{1}{(\delta\alpha_\mathrm{GA})^2}+\frac{1}{(\delta\bar\alpha)^2}\nonumber\\
\frac{\alpha_\mathrm{WA}}{(\delta\alpha_\mathrm{WA})^2}&=&\frac{\alpha_\mathrm{GA}}{(\delta\alpha_\mathrm{GA})^2}+\frac{\bar\alpha}{(\delta\bar\alpha)^2}\,,\label{eq:chisquarebariii}
\end{eqnarray}
effectively determining $\bar{\alpha}$ and $\delta\bar{\alpha}$.  With some redefinition of parameters, we can finally express the modified $\chi^2$ function (up to an irrelevant constant) as
\begin{eqnarray}
\Delta\chi^2({\bf z},{z}_{\alpha},1)&=&T^2\sum_{i=1}^N(\tilde{z}_i-\tilde{b}_i\tilde{z}_\alpha)^2+T_\alpha^2\tilde{z}_\alpha^2\nonumber\\
&\equiv&\Delta\chi^2(\tilde{\bf z},\tilde{z}_{\alpha})\,,\label{eq:chisquarebariv}
\end{eqnarray}
with

\begin{eqnarray}\label{eq:barpars}
\tilde{z}_{\alpha}&=&\frac{\alpha_s-\alpha_\mathrm{WA}}{\delta\alpha_\mathrm{WA}}\nonumber\\
\tilde{b}_{i}&=&(\delta\alpha_\mathrm{WA}/{\delta\alpha_\mathrm{GA}})b_i\\
\tilde{z}_i&=&z_i-b_i(\alpha_\mathrm{WA}-\alpha_\mathrm{GA})/\delta\alpha_\mathrm{GA}\nonumber\,.
\end{eqnarray}
Thus, with the redefined parameters, $\tilde{\bf z}$ and $\tilde{z}_\alpha$, the $\chi^2$ function is again identical to Eq.~(\ref{eq:chisquarebar}) from the appendix~\ref{sec:Dynamical Tolerances}, but now with $\alpha_s^0=\alpha_\mathrm{WA}$ and  $\delta\alpha_s=\delta\alpha_\mathrm{WA}$.  Thus, we see that our analysis of the combined PDF$+\alpha_s$ errors applies equally to either choice of $\alpha_s^0$ and $\delta\alpha_s$.

In practice, the penalty term in Eq.~(\ref{eq:chisquarebarii}) is not explicitly included in the global analysis ({\it i.e}., $k=0$ is used).  This has no effect on the extraction of the best-fit and error PDFs, since they are found by evaluating at fixed values of $\alpha_s$.  
Assuming that the $\chi^2$ function is exactly quadratic in $\alpha_s$, we can extract the tolerance parameter $T_\alpha^2$ from the values of $\chi^2$ at
the choices $\alpha_s=\alpha_\mathrm{WA}$ and $\alpha_s=\alpha_\mathrm{WA}\pm\delta\alpha_\mathrm{WA}$.  Letting
\begin{eqnarray}
\Delta^{\pm\alpha}&=&\Delta\chi^2(\alpha_\mathrm{WA}\pm\delta\alpha_\mathrm{WA})-\Delta\chi^2(\alpha_\mathrm{WA})\,,
\label{eq:barpars_deltachi2}
\end{eqnarray}
we obtain the result
\begin{eqnarray}
\frac{\Delta^{+\alpha}+\Delta^{-\alpha}}{2}&=&\left(\frac{\delta\alpha_\mathrm{WA}}{\delta\alpha_\mathrm{GA}}\right)^2(T^\alpha)^2\,.\nonumber
\end{eqnarray}
Note that this equation determines only the ratio $(T_\alpha/\delta\alpha_\mathrm{GA})$, since the $\alpha_s$ tolerance assumed in the global analysis and the resulting $\alpha_s$ uncertainty are inherently correlated.  However, assuming that the evaluation of the world average $\alpha_\mathrm{WA}$ is consistent with, and includes, the information from the global analysis, we expect $\delta\alpha_\mathrm{WA}/\delta\alpha_\mathrm{GA}<1$, which gives the requirement
\begin{eqnarray}
(T^\alpha)^2&>&\frac{\Delta^{+\alpha}+\Delta^{-\alpha}}{2}\,.\nonumber
\end{eqnarray}
In addition we also have the interesting asymmetry,
\begin{eqnarray}
\frac{\Delta^{+\alpha}-\Delta^{-\alpha}}{\Delta^{+\alpha}+\Delta^{-\alpha}}&=&\frac{2(\alpha_\mathrm{WA}-\alpha_\mathrm{GA})}{\delta\alpha_\mathrm{WA}}\,,\nonumber
\end{eqnarray}
which may be used to check the accuracy of the quadratic approximation in $\alpha_s$.

\section{Lagrange-Multiplier scans in \texttt{ePump}}\label{sec:LMScan}
The \texttt{ePump} code can perform the Lagrange-Multiplier scans efficiently. 
Using Eqs.~(\ref{eq:chisquare}) and Eq.~(\ref{eq:observableii}), the $\Delta \chi^2$ function supplemented by a Lagrange-Multiplier constraint can be written as:
\begin{eqnarray}
    \Delta\chi^2({\bf z},z_{\alpha})
    = T^2\left[\sum_{i=1}^N(z_i-b_iz_\alpha)^2+z_\alpha^2\right] + \lambda\left(\sum_{j=1}^{N+1}\Delta X^jz^\prime_j\right).
\end{eqnarray}
Because \texttt{ePump} updates the central values and uncertainties of the parameters $\{z_j^\prime\}$, the Lagrange-Multiplier constraint term is unchanged when the scan is applied to the updated PDF set.
The $\Delta\chi^2$ function after \texttt{ePump} updating, supplemented by the Lagrange-Multiplier constraint, takes the form:
\begin{eqnarray}
    \Delta\chi^2({\bf z},z_\alpha)_{\rm new}=T^2\sum_{i=1}^{N+1}(z^\prime_i)^2\,&+&\,\sum_{\alpha,\beta=1}^{N_X}\left(X_\alpha({\bf z},z_\alpha)-X_{\alpha}^E\right)C^{-1}_{\alpha\beta}\left(X_\beta({\bf z},z_\alpha)-X_{\beta}^E\right) \nonumber \\
    &+& \lambda\left(\sum_{j=1}^{N+1}\Delta X^jz^\prime_j\right)\,.
\end{eqnarray}
For the additional Lagrangian multiplier term, the $A^i$ and $M^{ij}$ should be
\begin{eqnarray}\label{eq:AM_LMScan}
A^i&=&
\frac{1}{-2T^2}\lambda\,\Delta{X}^{i}\,,\nonumber\\
M^{ij} &=&0,
\end{eqnarray} 
since the Lagrange-Multiplier constraint term contains no quadratic contribution.
These additional contributions to $A^i$ and $M^{ij}$ are added to the existing quantities used to construct the updating matrix.

\section{Correlation Ellipse between Observable and $\alpha_s$}\label{sec:Ellipse}

The correlation ellipse between two observables is the contour of the two-dimensional  $\Delta \chi^2$ surface, expressed as a function of those two observables; it can be derived using a Lagrange-Multiplier constraint.
To derive the correlation coefficient between an observable and $\alpha_s$, we supplement  Eqs.~(\ref{eq:chisquare}) and (\ref{eq:observableii}) with a Lagrange-Multiplier constraint:

\begin{eqnarray}
    \Delta\chi^2({\bf z},z_{\alpha})&=& T^2\left[\sum_{i=1}^N(z_i-b_iz_\alpha)^2+z_\alpha^2\right] + \lambda\left(\sigma(H)-\sigma(H_0)\right),\nonumber\\
    &=& T^2\left[\sum_{i=1}^N(z_i-b_iz_\alpha)^2+z_\alpha^2\right] + \lambda\left(\sum_{i=1}^N\Delta\sigma(H)_iz_i+\Delta\sigma(H)_\alpha z_\alpha\right),
\end{eqnarray}
where the $\sigma(H)$ is the $gg\rightarrow H$ cross section, taken here as a representative example of a general observable.
To locate the minimum of $\Delta \chi^2$, subject to the Lagrange-Multiplier constraint, we require the first derivatives to vanish:
\begin{eqnarray}
    \frac{\partial\Delta\chi^2}{\partial z_i}=2T^2(z_i-b_iz_\alpha)+\lambda\Delta\sigma(H)_i=0.
\end{eqnarray}
Solving for $z_i$ gives
\begin{eqnarray}
    z_i=-\frac{\lambda\Delta\sigma(H)_i}{2T^2}+b_iz_\alpha.
\end{eqnarray}
Summing over all PDF parameters, the Lagrange–Multiplier parameter $\lambda$ can be written as
\begin{eqnarray}
    \lambda=\dfrac{2T^2}{\displaystyle\sum_{i=1}^N\Delta\sigma(H)_i^2}\left[Bz_\alpha-(h-\Delta\sigma(H)_\alpha z_\alpha) \right],
\end{eqnarray}
where the $h$ denotes the $gg\rightarrow H$ cross section after subtracting its central prediction:
\begin{eqnarray}
    h=\sigma(H)-\sigma(H)_0=\sum_{i=1}^N \Delta\sigma(H)_iz_i+\Delta\sigma(H)_\alpha z_\alpha,
\end{eqnarray}
and $B$ is defined as 
\begin{eqnarray}
    B=\sum_{i=1}^N\Delta\sigma(H)_i b_i.
\end{eqnarray}
Substituting the expressions for  $z_i$ and $\lambda$, the minimized  $\Delta \chi^2$ of Eq.~(\ref{eq:chisquare}) takes the form
\begin{eqnarray}
    \Delta\chi^2_{\rm min}=T^2\dfrac{1}{\displaystyle \sum_{i=1}^N\Delta\sigma(H)_i^2}\left[Bz_\alpha-(h-\Delta\sigma(H)_\alpha z_\alpha)\right]^2+T^2z_\alpha^2.
\end{eqnarray}
Writing the above equation in matrix form:
\begin{eqnarray}
    \Delta\chi^2_{\rm min}=\begin{pmatrix}
     h, z_\alpha \\
    \end{pmatrix}\, T^2
    \begin{pmatrix}
    \dfrac{1}{\displaystyle \sum_{i=1}^N\Delta\sigma(H)_i^2} & -\dfrac{\Delta\sigma(H)_\alpha+B}{\displaystyle \sum_{i=1}^N\Delta\sigma(H)_i^2}  \\
    -\dfrac{\Delta\sigma(H)_\alpha+B}{\displaystyle \sum_{i=1}^N\Delta\sigma(H)_i^2} & 1+\dfrac{(\Delta\sigma(H)_\alpha+B)^2}{\displaystyle \sum_{i=1}^N\Delta\sigma(H)_i^2}  \\
    \end{pmatrix}\,
    \begin{pmatrix}
     h \\
     z_\alpha
    \end{pmatrix}\,.
\end{eqnarray}
The inverse matrix is
\begin{eqnarray}
    M^{-1}=\frac{1}{T^2}\begin{pmatrix}
        \displaystyle \sum_{i=1}^N\Delta\sigma(H)_i^2+(\Delta\sigma(H)_\alpha +B)^2 & \Delta\sigma(H)_\alpha +B \\
        \Delta\sigma(H)_\alpha +B & 1
    \end{pmatrix}.
\end{eqnarray}
The correlation coefficient is therefore 
\begin{eqnarray}
    \cos \phi=\dfrac{\Delta\sigma(H)_\alpha+\displaystyle \sum_{i=1}^N\Delta\sigma(H)_ib_i}{\sqrt{\displaystyle \sum_{i=1}^N\Delta\sigma(H)_i^2 + (\Delta\sigma(H)_\alpha+\displaystyle \sum_{i=1}^N\Delta\sigma(H)_ib_i)^2}}.
\end{eqnarray}
 
To eliminate the $b_i$ terms, we minimize $\Delta\chi^2$ in  Eq.~(\ref{eq:chisquare}) at $z_\alpha=\pm 1$.
When $z_\alpha=1$, after minimizing the $\Delta\chi^2$ function, $z_i=b_i$ will be obtained, while when $z_\alpha=-1$, $z_i=-b_i$ will be obtained. 
Accordingly, the $gg\rightarrow H$ cross sections evaluated with the 
$\alpha_s$ eigenvector PDF sets can be written as  

\begin{eqnarray}
    \sigma(H)(z_\alpha=1)&=&\sigma(H)_0+\sum_{i=1}^N\Delta\sigma(H)_ib_i+\Delta\sigma(H)_\alpha,\nonumber\\
    \sigma(H)(z_\alpha=-1)&=&\sigma(H)_0-\sum_{i=1}^N\Delta\sigma(H)_ib_i-\Delta\sigma(H)_\alpha,
\end{eqnarray}
so that the term containing $b_i$ can be expressed as
\begin{eqnarray}
    \Delta\sigma(H)_\alpha+\sum_{i=1}^N\Delta\sigma(H)_ib_i=\frac{\sigma(H)(z_\alpha=1)-\sigma(H)(z_\alpha=-1)}{2},
\end{eqnarray}
which is the $\alpha_s$-induced uncertainty on the $gg\rightarrow H$ cross section.
Finally, the correlation coefficient can be written as,
\begin{eqnarray}
    \cos \phi = \dfrac{\dfrac{\sigma(H)(z_\alpha=1)-\sigma(H)(z_\alpha=-1)}{2}}{\sqrt{\displaystyle\sum_{i=1}^N\Delta\sigma(H)_i^2 + (\dfrac{\sigma(H)(z_\alpha=1)-\sigma(H)(z_\alpha=-1)}{2})^2}},
\end{eqnarray}
which is the ratio of the $\alpha_s$-induced uncertainty to the quadrature sum of the PDF uncertainty and the $\alpha_s$-induced uncertainty.

\bibliography{ePump_alphaS_bib}

@article{ref:pdg,
    author = "Navas, S. and others",
    collaboration = "Particle Data Group",
    title = "{Review of particle physics}",
    doi = "10.1103/PhysRevD.110.030001",
    journal = "Phys. Rev. D",
    volume = "110",
    number = "3",
    pages = "030001",
    year = "2024"
}

@misc{ref:ATLAS_AlphaS,
    author = "Aad, Georges and others",
    collaboration = "ATLAS",
    title = "{A precise determination of the strong-coupling constant from the recoil of $Z$ bosons with the ATLAS experiment at $\sqrt{s} = 8$ TeV}",
    eprint = "2309.12986",
    archivePrefix = "arXiv",
    primaryClass = "hep-ex",
    month = "9",
    year = "2023"
}

@article{ref:CMS_AlphaS,
    author = "Chekhovsky, Vladimir and others",
    collaboration = "CMS",
    title = "{Determination of the strong coupling and its running from measurements of inclusive jet production}",
    eprint = "2412.16665",
    archivePrefix = "arXiv",
    primaryClass = "hep-ex",
    reportNumber = "CMS-SMP-24-007, CERN-EP-2024-327",
    doi = "10.1016/j.physletb.2025.139651",
    journal = "Phys. Lett. B",
    volume = "868",
    pages = "139651",
    year = "2025"
}

@article{ref:DIS_AlphaS,
    author = "Andreev, V. and others",
    collaboration = "H1",
    title = "{Determination of the strong coupling constant $\alpha_s(m_Z)$ in next-to-next-to-leading order QCD using H1 jet cross section measurements}",
    eprint = "1709.07251",
    archivePrefix = "arXiv",
    primaryClass = "hep-ex",
    reportNumber = "DESY-17-137, DESY17-137",
    doi = "10.1140/epjc/s10052-017-5314-7",
    journal = "Eur. Phys. J. C",
    volume = "77",
    number = "11",
    pages = "791",
    year = "2017",
    note = "[Erratum: Eur.Phys.J.C 81, 738 (2021)]"
}

@article{ref:FLAG,
    author = "Aoki, Y. and others",
    collaboration = "Flavour Lattice Averaging Group (FLAG)",
    title = "{FLAG review 2024}",
    eprint = "2411.04268",
    archivePrefix = "arXiv",
    primaryClass = "hep-lat",
    reportNumber = "CERN-TH-2024-192, FERMILAB-PUB-24-0785-T",
    doi = "10.1103/nfzp-p5dn",
    journal = "Phys. Rev. D",
    volume = "113",
    number = "1",
    pages = "014508",
    year = "2026"
}

@article{ref:Lattice_AlphaS,
    author = "d'Enterria, D. and others",
    title = "{The strong coupling constant: state of the art and the decade ahead}",
    eprint = "2203.08271",
    archivePrefix = "arXiv",
    primaryClass = "hep-ph",
    reportNumber = "FERMILAB-CONF-22-148-T",
    doi = "10.1088/1361-6471/ad1a78",
    journal = "J. Phys. G",
    volume = "51",
    number = "9",
    pages = "090501",
    year = "2024"
}

@article{ref:Reweighting_Hessian,
    author = "Paukkunen, Hannu and Zurita, Pia",
    title = "{Hessian PDF reweighting meets the Bayesian methods}",
    eprint = "1408.4572",
    archivePrefix = "arXiv",
    primaryClass = "hep-ph",
    doi = "10.22323/1.203.0048",
    journal = "PoS",
    volume = "DIS2014",
    pages = "048",
    year = "2014"
}

@article{ref:Reweighting_MC,
    author = "Ball, Richard D. and Bertone, Valerio and Cerutti, Francesco and Del Debbio, Luigi and Forte, Stefano and Guffanti, Alberto and Latorre, Jose I. and Rojo, Juan and Ubiali, Maria",
    collaboration = "NNPDF",
    title = "{Reweighting NNPDFs: the W lepton asymmetry}",
    eprint = "1012.0836",
    archivePrefix = "arXiv",
    primaryClass = "hep-ph",
    reportNumber = "EDINBURGH-2010-24, IFUM-968-FT, FR-PHENO-2010-040, RWTH-TTK-10-55",
    doi = "10.1016/j.nuclphysb.2011.03.017",
    journal = "Nucl. Phys. B",
    volume = "849",
    pages = "112--143",
    year = "2011",
    note = "[Erratum: Nucl.Phys.B 854, 926--927 (2012), Erratum: Nucl.Phys.B 855, 927--928 (2012)]"
}

@article{ref:ePump,
    author = "Schmidt, Carl and Pumplin, Jon and Yuan, C. -P.",
    title = "{Updating and optimizing error parton distribution function sets in the Hessian approach}",
    eprint = "1806.07950",
    archivePrefix = "arXiv",
    primaryClass = "hep-ph",
    reportNumber = "MSUHEP-18-006",
    doi = "10.1103/PhysRevD.98.094005",
    journal = "Phys. Rev. D",
    volume = "98",
    number = "9",
    pages = "094005",
    year = "2018"
}

@article{ref:ePump2,
    author = "Hou, Tie-Jiun and Yu, Zhite and Dulat, Sayipjamal and Schmidt, Carl and Yuan, C. -P.",
    title = "{Updating and optimizing error parton distribution function sets in the Hessian approach. II.}",
    eprint = "1907.12177",
    archivePrefix = "arXiv",
    primaryClass = "hep-ph",
    reportNumber = "MSUHEP-19-014",
    doi = "10.1103/PhysRevD.100.114024",
    journal = "Phys. Rev. D",
    volume = "100",
    number = "11",
    pages = "114024",
    year = "2019"
}

@article{ref:CT18,
    author = "Hou, Tie-Jiun and others",
    title = "{New CTEQ global analysis of quantum chromodynamics with high-precision data from the LHC}",
    eprint = "1912.10053",
    archivePrefix = "arXiv",
    primaryClass = "hep-ph",
    reportNumber = "MSUHEP-19-025, PITT-PACC-1911, SMU-HEP-19-03",
    doi = "10.1103/PhysRevD.103.014013",
    journal = "Phys. Rev. D",
    volume = "103",
    number = "1",
    pages = "014013",
    year = "2021"
}

@article{ref:E248,
    author = "Aaboud, Morad and others",
    collaboration = "ATLAS",
    title = "{Precision measurement and interpretation of inclusive $W^+$ , $W^-$ and $Z/\gamma ^*$ production cross sections with the ATLAS detector}",
    eprint = "1612.03016",
    archivePrefix = "arXiv",
    primaryClass = "hep-ex",
    reportNumber = "CERN-EP-2016-272",
    doi = "10.1140/epjc/s10052-017-4911-9",
    journal = "Eur. Phys. J. C",
    volume = "77",
    number = "6",
    pages = "367",
    year = "2017"
}

@article{ref:E268,
    author = "Aad, Georges and others",
    collaboration = "ATLAS",
    title = "{Measurement of the inclusive $W^\pm$ and Z/gamma cross sections in the electron and muon decay channels in $pp$ collisions at $\sqrt{s}=7$ TeV with the ATLAS detector}",
    eprint = "1109.5141",
    archivePrefix = "arXiv",
    primaryClass = "hep-ex",
    reportNumber = "CERN-PH-EP-2011-143",
    doi = "10.1103/PhysRevD.85.072004",
    journal = "Phys. Rev. D",
    volume = "85",
    pages = "072004",
    year = "2012"
}

@article{ref:E544,
    author = "Aad, Georges and others",
    collaboration = "ATLAS",
    title = "{Measurement of the inclusive jet cross-section in proton-proton collisions at $ \sqrt{s}=7 $ TeV using 4.5 fb$^{-1}$ of data with the ATLAS detector}",
    eprint = "1410.8857",
    archivePrefix = "arXiv",
    primaryClass = "hep-ex",
    reportNumber = "CERN-PH-EP-2014-155",
    doi = "10.1007/JHEP02(2015)153",
    journal = "JHEP",
    volume = "02",
    pages = "153",
    year = "2015",
    note = "[Erratum: JHEP 09, 141 (2015)]"
}

@article{ref:E218,
    author = "Aaij, R. and others",
    collaboration = "LHCb",
    title = "{Precision measurement of forward $Z$ boson production in proton-proton collisions at $\sqrt{s} = 13$ TeV}",
    eprint = "2112.07458",
    archivePrefix = "arXiv",
    primaryClass = "hep-ex",
    reportNumber = "LHCb-PAPER-2021-037, CERN-EP-2021-246",
    doi = "10.1007/JHEP07(2022)026",
    journal = "JHEP",
    volume = "07",
    pages = "026",
    year = "2022"
}

@article{ref:E553,
    author = "Aaboud, Morad and others",
    collaboration = "ATLAS",
    title = "{Measurement of the inclusive jet cross-sections in proton-proton collisions at $ \sqrt{s}=8 $ TeV with the ATLAS detector}",
    eprint = "1706.03192",
    archivePrefix = "arXiv",
    primaryClass = "hep-ex",
    reportNumber = "CERN-EP-2017-043",
    doi = "10.1007/JHEP09(2017)020",
    journal = "JHEP",
    volume = "09",
    pages = "020",
    year = "2017"
}

@article{ref:E554,
    author = "Aaboud, M. and others",
    collaboration = "ATLAS",
    title = "{Measurement of inclusive jet and dijet cross-sections in proton-proton collisions at $\sqrt{s}=13$ TeV with the ATLAS detector}",
    eprint = "1711.02692",
    archivePrefix = "arXiv",
    primaryClass = "hep-ex",
    reportNumber = "CERN-EP-2017-157",
    doi = "10.1007/JHEP05(2018)195",
    journal = "JHEP",
    volume = "05",
    pages = "195",
    year = "2018"
}

@article{ref:E581,
    author = "Tumasyan, Armen and others",
    collaboration = "CMS",
    title = "{Measurement of differential $t \bar t$ production cross sections in the full kinematic range using lepton+jets events from proton-proton collisions at $\sqrt {s}$ = 13{\,}{\,}TeV}",
    eprint = "2108.02803",
    archivePrefix = "arXiv",
    primaryClass = "hep-ex",
    reportNumber = "CMS-TOP-20-001, CERN-EP-2021-135",
    doi = "10.1103/PhysRevD.104.092013",
    journal = "Phys. Rev. D",
    volume = "104",
    number = "9",
    pages = "092013",
    year = "2021"
}

@article{ref:E528,
    author = "Sirunyan, Albert M and others",
    collaboration = "CMS",
    title = "{Measurements of $\mathrm{t\overline{t}}$ differential cross sections in proton-proton collisions at $\sqrt{s}=$ 13 TeV using events containing two leptons}",
    eprint = "1811.06625",
    archivePrefix = "arXiv",
    primaryClass = "hep-ex",
    reportNumber = "CMS-TOP-17-014, CERN-EP-2018-252",
    doi = "10.1007/JHEP02(2019)149",
    journal = "JHEP",
    volume = "02",
    pages = "149",
    year = "2019"
}

@article{ref:E545,
    author = "Khachatryan, Vardan and others",
    collaboration = "CMS",
    title = "{Measurement and QCD analysis of double-differential inclusive jet cross sections in pp collisions at $ \sqrt{s}=8 $ TeV and cross section ratios to 2.76 and 7 TeV}",
    eprint = "1609.05331",
    archivePrefix = "arXiv",
    primaryClass = "hep-ex",
    reportNumber = "CMS-SMP-14-001, CERN-EP-2016-196",
    doi = "10.1007/JHEP03(2017)156",
    journal = "JHEP",
    volume = "03",
    pages = "156",
    year = "2017"
}

@article{ref:ggHiggs1,
    author = "Ball, Richard D. and Bonvini, Marco and Forte, Stefano and Marzani, Simone and Ridolfi, Giovanni",
    title = "{Higgs production in gluon fusion beyond NNLO}",
    eprint = "1303.3590",
    archivePrefix = "arXiv",
    primaryClass = "hep-ph",
    reportNumber = "DCPT-13-30, DESY-13-001, EDINBURGH-2012-25, IFUM-1010-FT, IPPP-13-15",
    doi = "10.1016/j.nuclphysb.2013.06.012",
    journal = "Nucl. Phys. B",
    volume = "874",
    pages = "746--772",
    year = "2013"
}

@article{ref:ggHiggs2,
    author = "Bonvini, Marco and Ball, Richard D. and Forte, Stefano and Marzani, Simone and Ridolfi, Giovanni",
    title = "{Updated Higgs cross section at approximate N$^3$LO}",
    eprint = "1404.3204",
    archivePrefix = "arXiv",
    primaryClass = "hep-ph",
    reportNumber = "DESY-14-052, IFUM-1026-FT, DCPT-14-62, IPPP-14-31",
    doi = "10.1088/0954-3899/41/9/095002",
    journal = "J. Phys. G",
    volume = "41",
    pages = "095002",
    year = "2014"
}

@article{ref:ggHiggs3,
    author = "Bonvini, Marco and Marzani, Simone and Muselli, Claudio and Rottoli, Luca",
    title = "{On the Higgs cross section at N$^{3}$LO+N$^{3}$LL and its uncertainty}",
    eprint = "1603.08000",
    archivePrefix = "arXiv",
    primaryClass = "hep-ph",
    reportNumber = "OUTP-16-05P, TIF-UNIMI-2016-2",
    doi = "10.1007/JHEP08(2016)105",
    journal = "JHEP",
    volume = "08",
    pages = "105",
    year = "2016"
}

@article{ref:ggHiggs4,
    author = "Ahmed, Taushif and Bonvini, Marco and Kumar, M. C. and Mathews, Prakash and Rana, Narayan and Ravindran, V. and Rottoli, Luca",
    title = "{Pseudo-scalar Higgs boson production at N$^3$ LO$_{\text {A}}$ +N$^3$ LL $'$}",
    eprint = "1606.00837",
    archivePrefix = "arXiv",
    primaryClass = "hep-ph",
    reportNumber = "OUTP-16-13P",
    doi = "10.1140/epjc/s10052-016-4510-1",
    journal = "Eur. Phys. J. C",
    volume = "76",
    number = "12",
    pages = "663",
    year = "2016"
}

@article{ref:CTHiggsPaper,
    author = "Dulat, Sayipjamal and Hou, Tie-Jiun and Gao, Jun and Huston, Joey and Nadolsky, Pavel and Pumplin, Jon and Schmidt, Carl and Stump, Daniel and Yuan, C. -P.",
    title = "{Higgs Boson Cross Section from CTEQ-TEA Global Analysis}",
    eprint = "1310.7601",
    archivePrefix = "arXiv",
    primaryClass = "hep-ph",
    doi = "10.1103/PhysRevD.89.113002",
    journal = "Phys. Rev. D",
    volume = "89",
    number = "11",
    pages = "113002",
    year = "2014"
}

@misc{ref:CT25AS,
    author = "Ablat, Alim and Dulat, Sayipjamal and Guzzi, Marco and Huston, Joey and Mohan, Kirtimaan and Nadolsky, Pavel and Stump, Dan and Yuan, C. -P.",
    title = "{Strong Coupling Constant Determination from the new CTEQ-TEA Global QCD Analysis}",
    eprint = "2512.23792",
    archivePrefix = "arXiv",
    primaryClass = "hep-ph",
    reportNumber = "MSUHEP-25-019",
    month = "12",
    year = "2025"
}

@article{Zhan:2024tic,
    author = "Zhan, Wenxiao and Yang, Siqi and Liu, Minghui and Han, Liang and Stump, Daniel and Yuan, C. -P.",
    title = "{Improved Hessian method in global analysis of parton distribution functions}",
    eprint = "2411.11645",
    archivePrefix = "arXiv",
    primaryClass = "hep-ph",
    doi = "10.1103/858l-2x47",
    journal = "Phys. Rev. D",
    volume = "112",
    number = "7",
    pages = "074028",
    year = "2025"
}
\end{document}